\documentclass[twocolumn,english,preprintnumbers,amsmath,amssymb,pra,superscriptaddress,showpacs,unsrt]{revtex4}
\usepackage[T1]{fontenc}
\usepackage[utf8]{inputenc}
\usepackage{color}
\usepackage{graphicx}
\usepackage{amssymb}

\makeatletter

\providecommand{\tabularnewline}{\\}

\@ifundefined{textcolor}{}
{%
 \definecolor{BLACK}{gray}{0}
 \definecolor{WHITE}{gray}{1}
 \definecolor{RED}{rgb}{1,0,0}
 \definecolor{GREEN}{rgb}{0,1,0}
 \definecolor{BLUE}{rgb}{0,0,1}
 \definecolor{CYAN}{cmyk}{1,0,0,0}
 \definecolor{MAGENTA}{cmyk}{0,1,0,0}
 \definecolor{YELLOW}{cmyk}{0,0,1,0}
 }

\@ifundefined{definecolor}
 {\usepackage{color}}{}
\@ifundefined{definecolor}
 {\@ifundefined{definecolor}
 {\usepackage{color}}{}
}{}

\usepackage{dcolumn}

\usepackage{bm}

\usepackage{rotating}

\@ifundefined{definecolor}
 {\@ifundefined{definecolor}
 {\@ifundefined{definecolor}
 {\usepackage{color}}{}
}{}
}{}

\usepackage{subeqnarray}

\usepackage{psfrag}

\def\kbar{{\mathchar'26\mkern-9mu k}}
\def\pred{\mathfrak{p}}

\makeatother

\usepackage{babel}

\makeatother

\usepackage{babel}

\makeatother

\usepackage{babel}

\begin{document}

\title{Observation of the Anderson Metal-Insulator Transition with Atomic
Matter Waves: Theory and Experiment}

\author{Gabriel Lemari{\'e}}

\affiliation{Laboratoire Kastler Brossel, UPMC-Paris 6, ENS, CNRS; 4 Place Jussieu,
F-75005 Paris, France}

\author{Julien Chab{\'e}}

\altaffiliation[Present address: ]{Department of Physics, Ben-Gurion University P.O. Box 653 Be'er Sheva IL-84105 Israel }

\affiliation{Laboratoire PhLAM, Universit{\'e} de Lille 1, CNRS; CERLA; F-59655 Villeneuve d'Ascq Cedex, France}

\author{Pascal Szriftgiser}

\author{Jean Claude Garreau}

\affiliation{Laboratoire PhLAM, Universit{\'e} de Lille 1, CNRS; CERLA; F-59655 Villeneuve d'Ascq Cedex, France}

\author{Beno{\^i}t Gr{\'e}maud}

\affiliation{Laboratoire Kastler Brossel, UPMC-Paris 6, ENS, CNRS; 4 Place Jussieu,
F-75005 Paris, France}

\affiliation{Centre for Quantum Technologies, National University of Singapore,
3 Science Drive 2, Singapore 117543, Singapore}

\author{Dominique Delande}

\affiliation{Laboratoire Kastler Brossel, UPMC-Paris 6, ENS, CNRS; 4 Place Jussieu,
F-75005 Paris, France}

\date{\today}
\begin{abstract}
Using a cold atomic gas exposed to laser pulses -- a realization of
the chaotic quasiperiodic kicked rotor with three incommensurate frequencies
-- we study experimentally and theoretically the Anderson metal-insulator transition
in three dimensions. Sensitive measurements of the atomic wavefunction
and the use of finite-size scaling techniques make it possible to
unambiguously demonstrate the existence of a quantum phase transition
and to measure its critical exponents. By taking proper account of
systematic corrections to one-parameter scaling, we show the universality
of the critical exponent $\nu=1.59\pm0.01,$ which is found to be
equal to the one previously computed for the Anderson model.
\end{abstract}


\maketitle

\section{Introduction}

The interplay between quantum effects and disorder is a subject actively
studied for many decades, both theoretically and experimentally. It
plays a particularly important role in condensed matter physics, where,
in a first approximation, a crystal is modeled as independent electrons
interacting with a perfectly periodic lattice. The pioneering works
of Bloch and Zener \cite{Bloch:States:ZFP28,Zener:BlochOsc:PRSA34}
showed however that most predictions based on this model are not verified
in real crystals. For example, the Bloch theory predicts fully delocalized
wavefunctions implying a ballistic transport of the electrons through
the crystal. Moreover, in the presence of a constant bias potential,
Zener predicted an oscillatory motion (the Bloch-Zener oscillations)
due to quantum interference effects. This contradicts well-known experimental
facts at least in usual conditions.

An obvious possible explanation of these contradictions is the fact
that there are no perfect crystals: In a real crystal some sites may
be randomly occupied by ions of a different nature, thus breaking
the periodicity of the lattice. In 1958, Anderson considered this
approach and postulated that the dominant effect of the disorder is
to change randomly the on-site energy. Starting from this assumption,
he constructed a simple model \cite{Anderson:LocAnderson:PR58} of
a single-electron interacting with a lattice in the tight-binding
approximation: \begin{equation}
H_{\mathrm{tb}}=\sum_{jn}\epsilon_{jn}\vert jn\rangle\langle jn\vert+\sum_{jn,k\mu}V_{jn,k\mu}\vert jn\rangle\langle k\mu\vert\;.\label{eqAndersonH}\end{equation}
 Here $\epsilon_{jn}$ are the energies associated with the states
labeled by $n$ at the sites $j$ of the lattice, and the non-diagonal
elements $V_{jn,k\mu}$ denote the matrix elements between these
states. The diagonal part of the Hamiltonian corresponds to the potential
energy and the non-diagonal part to the kinetic energy in a continuous
space description. Disorder is introduced by giving the site energies $\epsilon_{jn}$
a \emph{random} distribution. Anderson thus showed that the electron
wavefunctions can be \emph{localized} by the disorder. This is naturally
in sharp contrast with the prediction of the Bloch model.

The phenomenon of localization has its most striking manifestation
in the transport properties of random media. If particle-particle
interactions are negligible, exponentially localized states cannot
contribute to transport at zero temperature since the coupling to
phonons is negligible. Anderson localization as a consequence of the
presence of disorder is one of the fundamental ingredients for the
understanding of the existence of insulators and metals, and, in particular,
the transition between the insulating and the metallic states of matter.
An insulator is associated with localized states of the system while
a metal generally displays diffusive transport associated with delocalized
states.

It was later shown that the 3D Anderson model displayed a phase
transition between a localized and a diffusive phase, the so-called
Anderson metal-insulator transition: If the disorder is below a critical
level, the localization disappears and one recovers a metallic (conductor)
behavior~\cite{Thouless:PREP74}. The link between
the disorder-induced metal-insulator transition and second-order phase-transitions
was established by reformulating the problem in terms of the renormalization
group \cite{Thouless:JPC72,Wegner:ZFP76}. Based on Wegner's work
and the ideas of Thouless and Landauer \cite{Landauer:PM70,Thouless:JPC72,Wegner:ZFP76},
it was possible to formulate the so-called \emph{one-parameter scaling
theory of localization} \cite{Abrahams:PRL79}, one of the most fruitful
approaches to the disorder-induced metal-insulator transition. The
essential hypothesis of the scaling theory is that, close to the transition,
a single relevant scaling variable describes the critical behavior.

An essential result of the one-parameter scaling theory is that the
Anderson transition exists only in dimensions larger than two. In
one dimension, \textit{all} electronic states are localized, whatever
\textit{\emph{the}} degree of randomness. In two dimensions, they
are all localized, but only marginally, i.e. with a localization length
exponentially large (thus possibly much larger than the sample size)
for weak disorder. 

In analogy to standard second-order phase-transitions,
the localization length $\ell$ is assumed to diverge at criticality
according to a power law: \begin{equation}
\ell\sim\left(W-W_{c}\right)^{-\nu}\;,\label{eq:Divergence}\end{equation}
with $\nu$ the localization length critical exponent, $W$ the disorder
strength and $W_{c}$ the critical disorder strength. The most important
assumption of the theory, the one-parameter scaling hypothesis, was
numerically validated using a finite-size scaling method developed
in \cite{MacKinnon:Kramer:PRL81,Pichard:Sarma:JPC81}. This technique,
which implements a real space renormalization, allowed to establish
numerically the existence of a scaling function for the localization
length. However, the critical exponents measured using this method,
$\nu\simeq1.57$ \cite{MacKinnon:JPC94,Slevin:PRL99}, were not compatible
with the result $\nu=1$ obtained from a self-consistent approach of localization
based on diagrammatic techniques, as developed in Ref.~\cite{Vollhardt:Wolfle:PRL82}.

In the half-century since its birth, the Anderson model has become
a paradigm for the studies of the interplay of quantum effects and
disorder. Despite that, relatively few experimental results are available,
for the following reasons: i) It is experimentally hard to finely
tune the disorder in a real crystal; ii) the decoherence sources (collision
with phonons, etc.) are difficult to master \cite{Lee:85}; iii) electrons
in a crystal present strong mutual interactions \cite{Altshuler:ANP06,Lee:85}
and iv) the wavefunction of the electrons in the crystal is not directly
accessible, only transport properties can be directly measured \cite{Kramer:Localization:RPP93}.

It is thus interesting to search for other systems
that display the Anderson transition, but are more favorable for experimental
studies. Indeed, the concept of Anderson localization has progressively
been extended from its original solid-state physics scope to a variety
of systems where a wave propagates in a disordered medium for example
electromagnetic radiation \cite{Maret:PRL06,Segev:N07} and sound waves
\cite{Condat:PRB87,Graham:PRL90,Page:NatPhys08}. Photons propagating in disordered
materials revealed to be an excellent system to observe the effects
of localization \cite{Maret:PRL06}. However, in such systems, there
is always some absorption, whose signature can be quite similar to
the signature of localization. Also, the measured quantity is the transmission,
and the wavefunction itself is not accessible. The recent experimental
observation of Anderson \textit{localization} \cite{Bouyer:AndersonBEC:N08}
using ultra-cold atomic matter waves has been done in a 1D situation
where states are always localized and no metal-insulator
\emph{transition} exists.

A very interesting Anderson-type system is obtained by combining
the Anderson model with another paradigmatic system, the kicked rotor
(KR), which has been theoretically studied for almost three decades.
This system is well known to be classically chaotic \cite{Chirikov:ChaosClassKR:PhysRep79},
and chaos plays here the role of a {}``dynamical'' disorder. In
the quantum case, the KR displays a localization phenomenon, called
{}``dynamical localization'' \cite{Casati:LocDynFirst:LNP79} which
is analogous to the 1D-Anderson localization \cite{Fishman:LocDynAnderson:PRA84}.
Moreover, a quasi-periodic generalization of the kicked rotor, substantially
equivalent to the 3D Anderson model, was numerically shown to display
an Anderson-like phase transition \cite{Casati:IncommFreqsQKR:PRL89}.
Experimental studies of the quantum kicked rotor were boosted by the
realization of such a system with laser-cooled atoms interacting with
a standing wave by Raizen and co-workers, which observed, for the
first time, the Anderson localization with matter waves \cite{Raizen:QKRFirst:PRL95}.

In the present paper we describe in detail a realization of an atomic
matter-wave system that allows us to observe the Anderson metal-insulator
transition \cite{AP:Anderson:PRL08}. We report a full characterization
of this phase transition which includes an experimental validation
of the one parameter scaling hypothesis and the first non ambiguous
experimental determination of the critical exponent $\nu$. Last but
not least, we show numerically that the quantum chaotic system we
consider has the \textit{same} critical behavior as the true random
3D Anderson model. In particular, we show that the two models belong
to the same universality class. Sec.~\ref{sec:KRStandard} introduces
the cold-atom realization of the periodic (standard) KR and its equivalence
with the 1D Anderson model, as well as the quasi-periodic generalization
of this system that is equivalent of the 3D-Anderson model. Sec.~\ref{sec:Experimentalrealization}
describes the corresponding experimental setup, paying attention to
its experimental limits (decoherence, stray effects, limited observation
time). In sec.~\ref{sec:crossover} we report our direct experimental
observation of the metal-insulator transition. In sec.~\ref{sec:CharacterizationPhaseTrans}
a scaling procedure is introduced that allows us to overcome experimental
limitations and determine the critical exponent corresponding to the
Anderson transition. Sec.~\ref{sec:nuetuniversality} is devoted
to the universality of the critical behavior. Sec.~\ref{sec:Conclusion}
concludes the paper.

\section{The atomic kicked rotor and its relation to the Anderson model\label{sec:KRStandard}}

\subsection{The atomic kicked rotor}

Consider a two level atom interacting with a laser standing wave of
frequency $\omega_{L}=k_{L}c$ detuned by $\Delta_{L}=\omega_{L}-\omega_{0}$
from the atomic transition of frequency $\omega_{0}$. It is well known that there are two kinds of interactions
between the atom and the radiation: Firstly, the atom can absorb a
photon from the laser and re-emit it spontaneously in a random direction.
This is a dissipative process giving rise to radiation pressure force,
whose rate is $\Gamma\Omega^{2}/4\Delta_{L}^{2}$ where $\Gamma$
is the natural width and $\Omega$ the resonant Rabi frequency (we
assume $|\Delta_{L}|\gg\Gamma$). Secondly, the atom can pick
a photon in a laser mode and emit it in the same (or another) laser
mode by stimulated emission. This conservative process is associated
with a potential acting on the atom's center of mass motion, called
the optical or dipole potential. For a standing wave this potential
is: \begin{equation}
V_{\mathrm{opt}}=\frac{\hbar\Omega^{2}}{8\Delta_{L}}\cos\left(2k_{L}X\right)\end{equation}
where $X$ is the atom center of mass position along the standing
wave. Clearly, this interaction is one dimensional, as momentum exchanges
between the atom and the radiation are always along the standing wave:
The atom absorbs a photon in one of the propagating beams and emits
it in the counterpropagating beam, leading to a quantized momentum
exchange of $2\hbar k_{L}$ along the $X$ axis. An important point
is that the optical potential amplitude scales as $\Omega^{2}/\Delta_{L}$
whereas the spontaneous emission rate scales as $\Gamma\Omega^{2}/\Delta_{L}^{2}$.
In the regime $|\Delta_{L}|\gg$$\Gamma,$ the optical potential is the
dominant contribution to the dynamics, with spontaneous emission events
being rare. Moreover, one can reduce the spontaneous emission rate
by increasing the detuning $\Delta_{L}$, provided that the laser
has enough power to keep the potential amplitude at the required level.

Suppose now that, instead of having the atom interacting continuously
with the standing wave, one modulates the radiation intensity periodically
(with period $T_{1}$) so that it is on for a short time $\tau$ (as
compared to the atom dynamics) and off the rest of the period. One
then obtains the Hamiltonian:

\begin{equation}
H=\frac{P^{2}}{2M}+\frac{\hbar\Omega^{2}\tau}{8\Delta_{L}}
\cos\left(2k_{L}X\right)\sum_{n}\delta_{\tau}(t^{\prime}-nT_{1})
\end{equation}
where $\delta_{\tau}(t)=1/\tau$ if $\vert t\vert\leq\tau/2$ and
zero otherwise. This functions tends to the Dirac $\delta$-function
as $\tau\rightarrow0$.

It is useful to introduce a set of scaled, dimensionless units: 
\begin{eqnarray}
\label{eq:scaling}
x & = & 2k_{L}X\nonumber \\
p & = & 2k_{L}T_{1}P/M\nonumber\\
t & = & t^{\prime}/T_{1}\nonumber\\
K & = & \frac{\hbar\Omega^{2}T_{1}\tau k_{L}^{2}}{2M\Delta_{L}}\\
\kbar & = & 4\hbar k_{L}^{2}T_{1}/M\nonumber\\
\mathcal{H} & = & \frac{4 k_L^2 T_1^2}{M} H\nonumber
\end{eqnarray}

In the limit of short pulses $\tau\ll T_{1}$, one then has: 
\begin{equation}
\mathcal{H}=\frac{p^{2}}{2}+K\cos x\sum_{n}\delta(t-n)\label{eq:HKR}
\end{equation}
which is precisely the Hamiltonian of the kicked rotor \cite{Chirikov:ChaosClassKR:PhysRep79,Izrailev:LocDyn:PREP90}.
One has thus realized an atomic kicked rotor \cite{Raizen:QKRFirst:PRL95}.
The above Hamiltonian is associated with the Schr\"odinger equation:
\begin{equation}
i\kbar\frac{\partial\psi}{\partial t}=\mathcal{H}\psi.
\end{equation}
$\kbar$ plays the crucial role of an effective Planck constant, which can
be adjusted at will by modifying e.g. the period $T_1$.
As shown in the following, the most interesting physics takes place
in the momentum. The scaling Eqs.~(\ref{eq:scaling}) is such
that $P=2\hbar k_L$ corresponds to $p=\kbar.$
If the atom is cold enough that its typical momentum is comparable
to $2\hbar k_{L}$ (the ``quantum'' of momentum exchange), quantum
effects can be observed in the system. 
Fortunately, magneto-optical
traps produce atoms with a typical momentum
of a few $\hbar k_{L}.$ It is customary to measure the atomic momentum $P$
in units of $2\hbar k_{L},$ i.e. measure $p$ in units of $\kbar.$
We thus will use:
\begin{equation}
\pred= \frac{p}{\kbar} = \frac{P}{2\hbar k_L}.
\end{equation}

For $K\gtrsim5,$ the classical KR is fully chaotic,
and the dynamics, although perfectly deterministic, behaves
like a pseudo-random diffusive process known as {}``chaotic diffusion''.
For this reason, $K$ is usually called ``stochasticity parameter''.
The existence of classical chaos can be seen by integrating the classical
equations of motion corresponding to Eq.~(\ref{eq:HKR}) over a period,
which leads to the so-called {}``Standard Map'': 
\begin{eqnarray}
x_{t+1}-x_{t} & = & p_{t}\\
p_{t+1}-p_{t} & = & K\sin x_{t+1}.
\end{eqnarray}

If the stochasticity parameter $K$ is large enough, $\sin x_{t}$
generates random numbers for successive $t$ values. The momentum
then performs a random (though deterministic) walk and the kinetic
energy (averaged over the initial conditions) increases linearly with
time. If -- as we assume in the following -- the initial state is
a narrow momentum distribution centered around the origin $\pred=0,$
one obtains: 
\begin{equation}
\langle \pred^{2}\rangle(t)=Dt\;,
\end{equation}
with $D\approx K^{2}/2\kbar^2$ being the diffusion constant.

In the quantum case, a chaotic diffusion is observed for times shorter
than a characteristic {}``localization time'' 
$\tau_{\mathrm{loc}}=D/2$,
after which quantum interferences build-up in the system that eventually
``freeze'' the dynamics, suppressing the diffusion. 
The mean kinetic
energy then tends to a constant 
$\langle \pred^{2}\rangle(t\rightarrow\infty)\rightarrow 2{\ell}^{2}$
with $\ell\approx K^{2}/4\kbar^2$ \cite{Shepelyansky:PRL86}. 
At the same time, the momentum distribution changes from a Gaussian
shape characteristic of a diffusive process to a localized, exponential
shape $\approx\exp\left(-|\pred|/{\ell}\right)$. This phenomenon is called
``dynamical localization'' (DL), ``dynamical'' meaning that
the localization takes place in momentum space. In fact, as shown
below, DL is intimately related to the Anderson localization, with,
however, an important difference: DL takes place
in \emph{momentum} space, whereas Anderson localization is in \emph{real}
space.

\subsection{Equivalence with the 1D-Anderson model}

Let us consider the KR quantum dynamics. From a stroboscopic point
of view, the motion is determined by the evolution operator over one
period: \begin{equation}
U=e^{-iK\cos x/\kbar}e^{-ip^{2}/2\kbar}\;,
\label{eq:U}
\end{equation}
 whose eigenstates form a basis set allowing to calculate the temporal
evolution. These {\it Floquet states} $\vert\phi\rangle$ are fully characterized
by their quasienergy $\omega$, defined modulo $2\pi$: 
\begin{equation}
U\vert\phi_{\omega}\rangle=e^{-i\omega}\vert\phi_{\omega}\rangle\;.
\label{eqQuasiStates}
\end{equation}

The Hamiltonian, Eq.~(\ref{eq:HKR}), is $2\pi$-periodic in position
$x,$ and so is the evolution operator, Eq.~(\ref{eq:U}). 
The Bloch theorem tells us that a Floquet eigenstate is a product
of a periodic function of $x$ by a plane wave 
$\exp{i\beta x}$ 
with $0\leq\beta<1,$ is a constant, $\beta\kbar$ being usually called
the ``quasi-momentum". A trivial transformation
shows that one can equivalently consider periodic functions of $x$
governed by the Hamiltonian, Eq.~(\ref{eq:HKR}), where $p$ 
is replaced
by $p+\beta\kbar.$ In the following discussion, we will omit for
simplicity the quasi-momentum, although it is straightforward to take
it into account. Note that in all numerical simulations shown hereafter, we
perform an averaging over the quasi-momentum, to follow the experimental
conditions where an incoherent sum of all quasi-momenta is prepared.

At this point, contact with a 1D Anderson tight-binding model can
be made by reformulating Eq.~(\ref{eqQuasiStates}) for the Floquet
states \cite{Fishman:LocDynAnderson:PRA84}. Firstly, we rewrite the (unitary) kick operator: 
\begin{equation}
e^{-iK\cos x/\kbar}=\frac{1+iW(x)}{1-iW(x)}\;,\end{equation}
 with 
 \begin{equation}
W(x)=\tan(K\cos{x}/2\kbar)\;.\label{eq:hoppingamplitudes3D}
\end{equation}
 The periodic function $W(x)$ 
 can be expanded in a Fourier series:
\begin{equation}
W(x)=\sum_{r}W_{r}e^{irx}.
\end{equation}
Similarly, for the kinetic part, one gets: \begin{equation}
e^{-i(p^{2}/2\kbar-\omega)}=\frac{1+iV}{1-iV}\;,
\end{equation}
 with $V$ diagonal in the momentum eigenbasis 
 $\vert m\rangle\equiv\vert p=\kbar m\rangle.$
Secondly, we make the following expansion in the momentum eigenbasis:
\begin{equation}
\frac{1}{1-iW(x)}\vert\phi_{\omega}\rangle=\sum_{m}\Phi_{m}\vert m\rangle\;.
\end{equation}
 Then, the eigen-equation for the Floquet state can be rewritten:
\begin{equation}
\epsilon_{m}\Phi_{m}+\sum_{r\neq0}W_{r}\Phi_{m-r}=-W_{0}\Phi_{m}\;,\label{eqAndersonmoedlKR}\end{equation}
 with $\epsilon_{m}=\tan\left[\frac{1}{2}(\omega-m^{2}\kbar/2)\right]$
 \cite{note:quasi-momentum}.

This is the equation for a tight-binding model with hopping elements
$W_{r}$ to the $r^\mathrm{th}$ neighbor, with eigen-energy $W_{0}$, and with
on-site energy $\epsilon_{m}$. The hopping elements are not restricted
to nearest-neighbors, but they decrease exponentially with $r$ \cite{note:hopping}.
In the original Anderson model, a random distribution is assigned
to $\epsilon_{m}$. Here, the sequence $\epsilon_{m}$, although not
satisfying the most stringent mathematical tests of randomness, is
nevertheless pseudo-random. These two conditions are sufficient for
the Anderson localization to take place. The hopping integrals $W_{r}$
increase with the kick strength $K$, which thus plays the role of
a control parameter in the Anderson model (\ref{eqAndersonmoedlKR}).
Note that if $\kbar$ is a rational multiple of $2\pi$, the $\epsilon_{m}$
are periodic in $m$. This leads to the quantum resonances of the kicked
rotor, where the states are extended.

When $\kbar$ is incommensurate with $2\pi$, the Floquet states are
found to be exponentially localized, and this property accounts for
dynamical localization. As shown in~\cite{AP:SubFMecs:EPL05}, 
the localization length observed at long times for a wavepacket
is essentially identical to the localization length of
individual Floquet states.  

Many references discuss the detailed correspondence between quantum
behavior of this dynamical system and Anderson localization: In Ref.
\cite{Izrailev:PhysRep90} an analogy between the KR and band random
matrices was pointed out; the latter have been reduced to a 1D nonlinear
$\sigma$ model \cite{Mirlin:PRL91} similar to those employed in
the localization theory \cite{Efetov:book97}. In Ref.~\cite{Atland:PRL96}
the direct correspondence between the KR and the diffusive supersymmetric
nonlinear $\sigma$ model was demonstrated. A diagrammatic approach
\cite{Wolfle:PRB80} to the dynamical localization in the Kicked Rotor
was reported in \cite{Atland:PRL93}.

\subsection{The quasi-periodic Kicked Rotor and its analogy to the 3D-Anderson
model}

\label{sec:KRquasiper}

As the Anderson transition exists only in three (or more) dimensions,
one must generalize the KR to obtain a system analogous to a 3D Anderson
model.

Different generalizations of the KR have been theoretically considered
as analogs of the 3D-Anderson model \cite{Fishman:PRL88,GarciaGarcia:PRE09}.
Here we use the convenient three-incommensurate-frequencies generalization
introduced in Refs.~\cite{Shepelyansky:PRL89,Shepelyansky:PD83}: \begin{equation}
\mathcal{H}_{\mathrm{qp}}=\frac{p^{2}}{2}+\mathcal{K}(t)\cos x\sum_{n}\delta(t-n)\;,\label{eq:KRquasiper}\end{equation}
 obtained simply by modulating the amplitude of the standing wave
pulses with two new frequencies $\omega_{2}$ and $\omega_{3}$: 
\begin{equation}
\mathcal{K}(t)=K\left[1+\varepsilon\cos\left(\omega_{2}t+\varphi_{2}\right)\cos\left(\omega_{3}t+\varphi_{3}\right)\right]\;.\label{eq:Kdet}
\end{equation}
 One can legitimately ask: where is the three dimensional aspect in
the latter Hamiltonian? An answer can be given by drawing a formal
analogy between the quasiperiodic kicked rotor and a 3D kicked rotor
with an initial condition taken as ``plane source'' (see below).

We start from the Hamiltonian of a 3D periodically kicked rotor: 
\begin{eqnarray}
\lefteqn{\mathcal{H}=\frac{p_{1}^{2}}{2}+\omega_{2}p_{2}+\omega_{3}p_{3}}\nonumber \\
 &  & +K\cos x_{1}\left[1+\varepsilon\cos x_{2}\cos x_{3}\right]\sum_{n}\delta(t-n)\;,
 \label{eqKR3DquasiperH}
 \end{eqnarray}
let us consider the evolution of a wavefunction $\Psi$ with the initial condition: 
\begin{equation}
\label{eqPsi3}
 \Psi({x}_{1},{x}_{2},{x}_{3},t=0)\equiv\Xi({x}_{1},t=0)\delta({x}_{2}-\varphi_{2})\delta({x}_{3}-\varphi_{3})
\end{equation}

 The initial state being perfectly localized in $x_{2}$ and $x_{3}$,
it is entirely delocalized in the conjugate momenta $p_{2}$ and $p_{3}$,
and can thus be seen as a ``plane source" \cite{Weaver:PRE05}
in  momentum space.

From a stroboscopic point of view, the time-evolution of $\Psi$ is
determined by the evolution operator over one period: 
 \begin{equation}
\label{eq:UKR3D}
\mathcal{U}=e^{-iK\cos x_{1}(1+\varepsilon\cos{x}_{2}\cos{x}_{3})/\kbar} \times e^{-i\left(p_{1}^{2}/2+\omega_{2}p_{2}+\omega_{3}p_{3}\right)/\kbar}.
\end{equation}
It is then straightforward to see that the 3D-wave function $\Psi$
at time $t$ is related to its initial condition as: 
\begin{eqnarray}
\lefteqn{\Psi({x}_{1},{x}_{2},{x}_{3},t) = \mathcal{U}^{t}\Psi({x}_{1},{x}_{2},{x}_{3},t=0)}\nonumber \\
 & = & \Xi({x}_{1},t)\delta({x}_{2}-\varphi_{2}-\omega_{2}t)\delta({x}_{3}-\varphi_{3}-\omega_{3}t)\;,\label{eq:3d}
\end{eqnarray}
 with: \begin{widetext}\begin{equation}
\Xi({x}_{1},t)\equiv\prod_{t'=1}^{t}\; e^{-iK\cos{x}_{1}[1+\varepsilon\cos(\varphi_{2}+\omega_{2}t')\cos(\varphi_{3}+\omega_{3}t')]/\kbar}e^{-ip_{1}^{2}/2\kbar}\;\Xi({x}_{1},t=0)\;.\label{eqevolopKRpsit}\end{equation}
 \end{widetext}

On the other hand, consider now the evolution of an initial wave function
$\psi({x},t=0)$ with the Hamiltonian $H_{\mathrm{qp}}$ of the quasiperiodic
kicked rotor. It is also determined by an evolution operator from
kick to kick, but now this evolution operator $U_{\mathrm{qp}}(t;t-1)$
depends on time, since the Hamiltonian $H_{\mathrm{qp}}$, Eq.~(\ref{eq:KRquasiper}),
is not time-periodic: \begin{eqnarray}
\lefteqn{U_{\mathrm{qp}}(t;t-1)=}\nonumber \\
 &  & e^{-iK\cos{x}[1+\varepsilon\cos(\varphi_{2}+\omega_{2}t)\cos(\varphi_{3}+\omega_{3}t)]/\kbar}e^{-i{p}^{2}/2\kbar}\;.\label{eqevolopKRquasiper}\end{eqnarray}
 The wave-function $\psi(t)$ at time $t$ is obtained by applying
successively $U_{\mathrm{qp}}(t';t'-1)$ for $t'$ from $1$ to $t$:
\begin{equation}
\psi({x},t)=\prod_{t'=1}^{t}U_{\mathrm{qp}}(t';t'-1)\psi({x},t=0)\;.\label{eqevolKRquasiper}\end{equation}

From Eqs.~(\ref{eq:3d}), (\ref{eqevolopKRquasiper}), (\ref{eqevolopKRpsit})
and (\ref{eqevolKRquasiper}), it follows that $\psi({x},t)$ and
$\Xi(x_{1},t)$ follow exactly the same evolution. Consequently, the
dynamics of the quasiperiodic kicked rotor is \emph{strictly} equivalent
to that of a 3D kicked rotor with a plane source. Our experiment with
the quasiperiodic kicked rotor can be seen as a localization
experiment in a 3D disordered system, where localization is
actually observed in the direction perpendicular
to the plane source \cite{Page:NatPhys08}. In other words, the situation is thus comparable to a transmission experiment
where the sample is illuminated by a plane wave and the exponential localization is only measured along the 
wave vector direction. Therefore, the behavior of the wave function $\psi$ subjected
to the quasiperiodic kicked rotor Hamiltonian $H_{\mathrm{qp}}$,
Eq.~(\ref{eq:KRquasiper}), depicts \textit{all} the properties of
the dynamics of the quantum 3D kicked rotor, Eq.~(\ref{eqKR3DquasiperH}).

The Hamiltonian $\mathcal{H}$, Eq.~(\ref{eqKR3DquasiperH}), is
invariant under the following transformation, product of time-reversal
with parity: \begin{equation}
T:t\rightarrow-t,\mathbf{x}\rightarrow-\mathbf{x},\mathbf{p}\rightarrow\mathbf{p}\;,\label{eq:TRS}\end{equation}
which is relevant for dynamical localization \cite{Scharf:JPA89,Smilansky:PRL92}.
The evolution of the states according to the Hamiltonian, Eq.~(\ref{eq:KRquasiper}),
is governed by the operator $\mathcal{U}$, Eq.~(\ref{eq:UKR3D}), which belongs to
the Circular Orthogonal Ensemble class \cite{HaakeQSC,StoeckmannChaos},
with the additional constraint at $t=0$ Eq.~(\ref{eqPsi3}). 
Of course, the transformation~\eqref{eq:TRS} 
amounts to changing $(\varphi_{2},\varphi_{3})$ to $(-\varphi_{2},-\varphi_{3})$ into the constraint~\eqref{eqPsi3}, 
i.e. to starting from a different wavefunction. On the other hand, from Eq.~\eqref{eq:3d}, one clearly sees that after 
$t$ steps, the constraint reads $\delta({x}_{2}-\tilde{\varphi_{2}})\delta({x}_{3}-\tilde{\varphi_{3}})$, with
\begin{equation}
 \tilde{\varphi_{2}}=\varphi_{2}+\omega_{2}t\qquad \tilde{\varphi_{3}}=\varphi_{3}+\omega_{3}t.
\end{equation}
Since the frequencies $\omega_2$ and $\omega_3$ are incommensurate, the preceding equation 
immediately tells us that, along the time evolution, the constraint on the wavefunction can be arbitrary close to any
phases $(\varphi_{2}',\varphi_{3}')$~\cite{Basko:PRL03}. This way, the time evolution results in an average over (almost) all possible phases, showing thus that
the localization properties are independent of a particular choice $(\varphi_{2},\varphi_{3})$, 
but only depend on the operator $\mathcal{U}$. Therefore, 
the dynamical properties of the present quasiperiodic kicked
rotor also belong to the orthogonal ensemble.

It should be noted that the 3D aspect comes from the presence of 3 frequencies
in the dynamical system: the usual ``momentum frequency''
$\kbar$ present in the standard kicked rotor Eq.~(\ref{eq:HKR}),
and two additional time-frequencies $\omega_{2}$ and $\omega_{3}$.
Thus, increasing the number of incommensurate frequencies allows one
to tune the effective dimensionality of the system.

Let us now consider the conditions for the observation of Anderson
localization with the quasiperiodic kicked rotor. As for the
standard kicked rotor, the Floquet states of the
time-periodic 3D Hamiltonian $\mathcal{H}$, Eq.~(\ref{eqKR3DquasiperH}),
can be mapped onto a 3D Anderson-like model: 
\begin{equation}
\epsilon_{\mathbf{m}}\Phi_{\mathbf{m}}+\sum_{\mathbf{r}\neq0}W_{\mathbf{r}}\Phi_{\mathbf{m}-\mathbf{r}}=-W_{\mathbf{0}}\Phi_{\mathbf{m}}\;,
\label{eqAndersonmodelKRquasiper}
\end{equation}
 where $\mathbf{m}\equiv(m_{1},m_{2},m_{3})$ and $\mathbf{r}$ label
sites in a 3D cubic lattice, the on-site energy $\epsilon_{\mathbf{m}}$
is 
\begin{equation}
\epsilon_{\mathbf{m}}=\tan\left\lbrace \frac{1}{2}\left[\omega-\left(\kbar\frac{{m_{1}}^{2}}{2}+\omega_{2}m_{2}+\omega_{3}m_{3}\right)\right]\right\rbrace \;,
\label{eq:pseudo-random-disorder-quasiperKR}
\end{equation}
 and the hopping amplitudes $W_{\mathbf{r}}$ are coefficients of
a threefold Fourier expansion of 
\begin{equation}
W({x}_{1},{x}_{2},{x}_{3})=\tan\left[K\cos{x}_{1}(1+\varepsilon\cos{x}_{2}\cos{x}_{3})\right/2\kbar]\;.
\label{eq:hoppingamplitudes3DQKR}
\end{equation}

\begin{figure}
\begin{centering}
\psfrag{Momentum}{Momentum $p$ } 
\includegraphics[width=7cm]{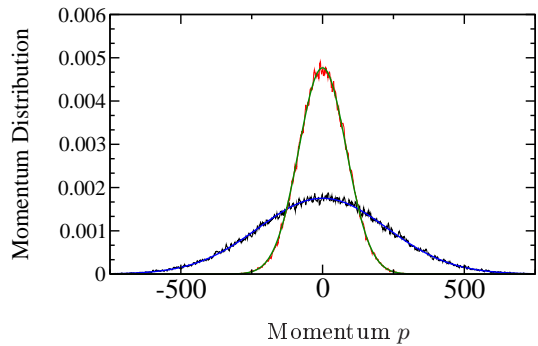} 
\end{centering}
\caption{\label{fig:classical_diffusion_3D}
(Color on line) Classical diffusive motion for the 3D kicked rotor Eq.~\ref{eqKR3DquasiperH}. 
The initial state is localized around
the origin. After 1000 kicks, the classical momentum distribution
(black and red curves) 
has the Gaussian shape characteristic of a diffusive motion. The blue
and green curves are fits by a Gaussian which do not show any statistically
significant deviation. The black (resp. red) curve is the momentum distribution
along $p_1$ (resp. $p_2$). The distribution along $p_3$ is identical that
along $p_2.$ The anisotropic diffusion happens because the 
hopping along
the directions "2" and "3" is  diminished by a factor $\varepsilon$ compared
to hopping along direction "1". Parameters are $K=10$, $\kbar=2.85$, $\varepsilon=0.8$,
$\omega_2/2\pi=\sqrt{5}$, $\omega_3/2\pi=\sqrt{13}.$}
\end{figure}

An obvious necessary condition for the observation of localization
effects is that $\epsilon_{\mathbf{m}}$ is not periodic. 
This is achieved if $(\kbar,\omega_{2},\omega_{3},\pi)$
are incommensurate. Of course, the presence of \textit{disorder}
in the diagonal energy $\epsilon_{\mathbf{m}}$ is crucial to observe
Anderson localization. When $\kbar$ is incommensurate with $2\pi$,
due to the presence of a nonlinear dispersion in the $m_{1}$ direction,
the classical dynamics can become chaotic with diffusive spreading
in \textit{all} $\mathbf{m}$ directions \cite{Shepelyansky:PD97,Shepelyansky:PRL89}.
A typical numerical simulation is shown in Fig.~\ref{fig:classical_diffusion_3D}: 
the classical motion is almost perfectly diffusive along the
three $p_i$ coordinates with a characteristic Gaussian shape in each
direction. From Eq.~(\ref{eq:hoppingamplitudes3DQKR}, it is clear that hopping along
the directions "2" and "3" is  diminished by a factor $\varepsilon$ compared
to hopping along direction "1". Not surprisingly, diffusion along $p_1$
is slightly faster than along $p_2$ and $p_3.$ The quasi-periodically kicked
rotor is thus analogous to an anisotropic Anderson 
model~\cite{Soukoulis:PRB89,Evangelou:PRB94,Soukoulis:PRL96}.

When those conditions are verified, localization effects as predicted
for the 3D Anderson model are expected, namely either a diffusive
or a localized regime. Localized states would be observed if the disorder
strength is large as compared to the hopping. In the case of the model
Eq.~(\ref{eqAndersonmodelKRquasiper}), the amplitude of the disorder
is fixed, but the hopping amplitudes can be controlled by changing
the stochasticity parameter $K$ (and/or the modulation amplitude
$\varepsilon$): $W_{\mathbf{r}}$ is easily seen to increase with
$K$. In other words, the larger $K$, the smaller the disorder. One
thus expects to observe diffusive regime for large stochasticity or/and
modulation amplitude (small disorder) and localized regime for small
$K$ or/and $\varepsilon$ (large disorder). It should be emphasized
that there is no \emph{stricto sensu} mobility edge in our system.
Depending on the values of the parameters $K,\kbar,\varepsilon,\omega_{2},\omega_{3},$
all Floquet states are localized or all are delocalized. The boundary
of the metal-insulator transition is in the $K,\kbar,\varepsilon,\omega_{2},\omega_{3}$
space. As seen below, $K$ and $\varepsilon$ are the primarily important
parameters.

In an analytical work on a similar problem \cite{Basko:PRL03} 
Basko \emph{et al.} showed that the weak dynamical localization regime of a $d$-frequency
quantum dot system is similar to the weak localization
in a $d$-dimensional Anderson model. This work confirms the equivalence
between our system and the 3D-Anderson model. The above arguments
were also validated numerically \cite{Shepelyansky:PD97,Shepelyansky:PRL89}.

Numerical simulations of the evolution of the quasi-periodically kicked
rotor are straightforward. The free evolution between consecutive
pulses is diagonal in momentum representation, while the kick operator
is diagonal in position representation (whatever the kick strength,
constant or quasi-periodic). Switching between momentum and position
representation is easily done through a Fast Fourier Transform. We
are thus able to compute the evolution of a large number of initial
states (typically one thousand) over a very long time (typically up
to one million kicks, much more than in the experiment). All numerical
results shown below have been carefully checked for convergence. Except
when explicitly stated, averaging over the quasi-momentum $\beta$
has been performed, in accordance with the experimental realization.

\section{Experimental realization with atomic matter-waves\label{sec:Experimentalrealization}}

\subsection{Experimental setup}

Our experimental setup has been described in detail in previous publications
\cite{AP:DiodeMod:EPJD99,AP:RamanSpectro:PRA01,AP:ChaosQTransp:CNSNS:2003,AP:Polarization:OC07}
and was used in various investigations on the quasiperiodic kicked
rotor \cite{AP:Bicolor:PRL00,AP:SubFourier:PRL02,AP:Reversibility:PRL05,AP:PetitPic:PRL06}.
Briefly, our experiments are performed with cesium atoms produced
in a standard magneto-optical trap (MOT). A long Sisyphus-molasses
phase (25 ms) allows us to obtain $10^{7}$ atoms at a measured temperature
of $3.2$ $\mu$K. The velocity distribution of the atoms is well
modeled by an incoherent sum of plane waves forming a Gaussian of
full width at half maximum (FWHM) equal to $8\hbar k_{L}$, which
is much narrower than the expected localization length. The MOT beams
and magnetic field are turned off and the sequence of kicks is applied
to the atoms. The beam forming the standing wave passes through an
acousto-optical modulator driven by RF pulse synthesizers, which generates
the kicks at a typical frequency of $1/T_{1}=36$ kHz (which corresponds
to $\kbar=2.89$), of duration $\tau=$900 ns and with a raising
time of 50 ns. The modulation is thus an almost perfect square, at
the time scale of the atomic motion, and its duration and period can
be set by a microcomputer. The beam is then injected in an optical
fiber that brings it to the interaction region, and the standing wave
is obtained simply by back-reflection of this beam. The standing wave has
a typical power 160 mW, its profile intensity has a
FWHM of 1.5 mm, and it is far off-resonant (7.3 GHz to red, or $1.4\times10^{3}\Gamma$),
in order to reduce spontaneous emission. The corresponding stochasticity
parameter is $K\approx15$.

A very interesting property of our system (as compared to solid-state
systems) is that the wave function is accessible (or at least its
square modulus). We measure the atomic velocity distribution by velocity-selective
Raman stimulated transitions, which are sensitive to the atomic velocity
via Doppler effect, allowing an optimal velocity resolution of about
2 mm/s. A Raman pulse detuned of $\delta_{R}$ with respect to the
Raman resonance transfers the atoms in the velocity class $v=\delta_{R}/(2k_{R})-v_{R}$
with $v_{R}=\hbar k_{R}/M$ ($k_{R}$ is the wave number of the Raman
beams) from the $F_{g}=4$ to the $F_{g}=3$ ground-state hyperfine
sublevel. A beam resonant with the $F_{g}=4\rightarrow F_{e}=5$ transition
is then applied to push the remaining atoms out of the interaction
region. The $F_{g}=3$ atoms are then optically pumped to the $F_{g}=4$
sublevel and interact with a resonant probe beam: The absorption signal
is thus proportional to the population of the $F_{g}=4$ level, thus
to the population of the selected velocity class. The whole sequence
then starts again with a different value of the Raman detuning to
probe a new velocity class, allowing a reconstruction of the velocity
distribution \cite{AP:RamanSpectro:PRA01,AP:Polarization:OC07}.

\subsection{Decoherence sources}

Any quantum experiment must consider decoherence sources that destroy
quantum interference effects (in our case, localization) reestablishing
a diffusive dynamics. The most important sources of decoherence in
our experiment are (i) atomic collisions, (ii) spontaneous emission,
and (iii) the deviation of the standing wave from strict horizontality.

For an isolated system described by a single wavefunction, phase coherence
between different positions is ``perfect". When the
system is weakly coupled to an external bath, it cannot be any longer
described by a single wavefunction; the most convenient description
usually involves a density matrix $\rho.$ Non-diagonal matrix elements
of the type $\langle x|\rho|x'\rangle$ quantify the degree of coherence
of the system between position $x$ and $x'.$ As a general rule,
the effect of the external bath is to make the non-diagonal elements
of the density matrix to decay relatively rapidly, more rapidly than
the diagonal elements: this is decoherence (not to be confused with
dissipation)~\cite{Zurek:RMP03}. Effects like Anderson localization are due to subtle destructive
interference amongst various components of the wavefunction, which
inhibit the classically allowed transport: they are thus very sensitive
to decoherence. One usually quantify the strength of decoherence effects
by defining a phase coherence time, the characteristic time over which the non-diagonal
elements of the density matrix decay because of coupling to the external
bath. In our case, the non-diagonal element of interest are between
eigenstates $|p\rangle$ and $|p'\rangle$ located at a typical distance
$|p-p'|$ comparable to the localization length in momentum space.

Localization effects can be observed only for times shorter than the
phase coherence time \cite{Janssen:PREP98}. Beyond the phase coherence
time, interference effects are killed and classical-like diffusive
dynamics sets in. In the following, we shall express the characteristic
times of the decoherence processes (i), (ii) and (iii), as functions
of the experimental parameters to show that they can be set large
enough for localization effects to be observable.

In atom-atom collisions, the dominant effect is that of collisions
between cold atoms, the density of the cloud being around 8 orders
of magnitude larger than the density of the background hot gas. A
cloud density of $10^{12}$ cm$^{-3}$ with a mean velocity 1 cm/s
and a collision cross-section of $6\times10^{-11}$ cm$^{2}$ gives
a collision rate of $\approx60$ s$^{-1}$, or $1.6\times10^{-3}$
per kick; the collision phase coherence time is thus $\sim600$ kicks.

In order to have a better idea of the decoherence effect induced by
spontaneous emission, let us consider the temporal evolution of an
initial plane-wave function: $\psi(p,t=0)=\delta(p-p_{0})$ evolving
with the KR Hamiltonian Eq.~(\ref{eq:HKR}). After dynamical localization
sets in, the momentum distribution ceases to expand because
of destructive interference between the various components
of the wavefunction. Spontaneous emission brings a random
recoil to the atomic momentum which is not an integer multiple of
$2\hbar k_L.$ Thus, the quasi-momentum $\beta$ performs a random jump.
As the phase factors involved in the free evolution depend on
the quasi-momentum, the relative phases between interfering paths
are scrambled, resulting in a  new transient diffusive behavior
for another duration of $\tau_{\mathrm{loc}}$. 
DL is thus expected to be destroyed if spontaneous emission is regularly
repeated. Note that a single spontaneous emission event completely
breaks the phase coherence, implying that the phase coherence time
is simply the inverse of the spontaneous emission rate.

Spontaneous emission tends to reestablish a
diffusive evolution with a diffusion constant that is roughly $\eta\kbar^{2}$
where $\eta=\Gamma\Omega^{2}\tau/8\Delta_{L}^{2}$, is the spontaneous
emission rate expressed in photons per kick, which can be cast in
the more useful form $\eta=\left(\Gamma\tau/8\right)\left(I/I_{s}\right)\left(\Gamma/\Delta_{L}\right)^{2}$,
where $I$ is the intensity and $I_{s}\approx2.2$ mW{/cm}$^{2}$
is the transition saturation intensity. Around the transition $(K\approx6)$,
the experimental values indicated above give $\eta\approx2.1\times10^{-3}$
s$^{-1}$, or a typical phase coherence time of $\sim$500 kicks.

\begin{figure}
\begin{centering}
\psfrag{x}{$t$ } 
\psfrag{y}{$\langle \pred^{2}\rangle$ } 
\includegraphics[width=7cm]{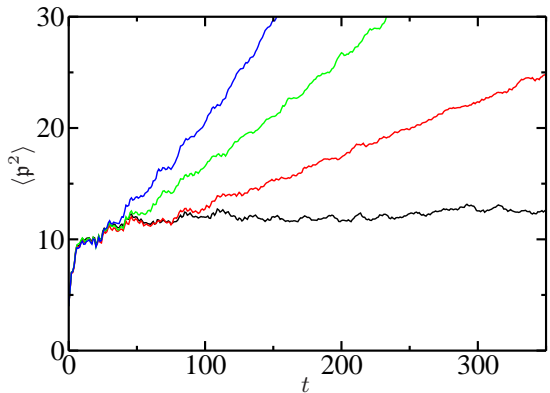} 
\par\end{centering}

\caption{\label{fig:simup2vstvsinclin}(Color online) Gravity effects on a
slightly inclined kicked rotor Eq.~(\ref{eq:HKRgravity}). The deviation
of the standing wave from horizontality is $\alpha=0^{\circ}$ (black
lower curve), $\alpha=0.1^{\circ}$ (red lower middle curve), $\alpha=0.4^{\circ}$
(green upper middle curve) and $\alpha=1^{\circ}$ (blue upper curve).
The stochasticity parameter is taken as $K=5$ and the effective Planck
constant is $\kbar=2.85$. The dynamics of an initial thermal state
is simulated and the corresponding mean kinetic energy is plotted
versus time. For angles larger than $0.1^{\circ}$, the slow drift
of momentum induces a diffusive behavior clearly visible on the time-scale
of the experiment.}

\end{figure}

Another effect leading to the destruction of localization is the standing
wave deviation from horizontality. In this case, a gravity term must
be added in the Hamiltonian (\ref{eq:HKR}): 
\begin{equation}
H_{g}=\frac{p^{2}}{2}-\eta_{g}x+K\cos x\sum_{n}\delta(t-n)\;,\label{eq:HKRgravity}
\end{equation}
 The dimensionless gravity term $\eta_{g}$ is: 
 \begin{eqnarray}
\eta_{g} & = & \frac{mgT_{1}}{2\hbar k_{L}}\;\kbar\;\sin\alpha\;,
\end{eqnarray}
 with $g$ the gravity acceleration and $\alpha$ the angle between
the horizontal direction and the standing wave. The physical interpretation
is quite clear: $mgT_{1}\sin\alpha$ is the additional momentum transferred
to the atoms between two consecutive kicks, which must be compared
to the width of the Brillouin zone $2\hbar k_{L}.$

The gravity term $-\eta_{g}x$ breaks the spatial periodicity of the
Hamiltonian, and consequently the conservation of the quasi-momentum 
$\beta\kbar.$ It actually produces a drift of the quasi-momentum
at constant rate $-\eta_{g},$ whose effect is to break dynamical 
localization. Indeed, the destructive interference between 
various components of the momentum wavefunction -- responsible
for dynamical localization -- is partially destroyed by the
quasi-momentum drift, as the various phase factors accumulated
during the free evolution between two consecutive kicks,
$\exp\left[-i(m+\beta\kbar)^2/2\kbar\right]$ 
also drift. The net result is a residual
diffusion constant, depending on $\eta_{g}.$ Although this is not
strictly a decoherence effect (the whole evolution is fully phase
coherent), it similarly destroys dynamical localization.
We thus define the phase coherence time $\tau_{g}$ as the time
needed to double $\langle p^2 \rangle$ compared to the
dynamically localized situation. Numerical simulations taking
into account the gravity effect confirm the discussion above, see
Fig.~\ref{fig:simup2vstvsinclin}. If the standing wave deviates
from horizontality by an angle $\alpha=1^{\circ}$, then $\tau_{g}\approx120$
kicks whereas when the angle $\alpha=0.1^{\circ}$, $\tau_{g}\approx350$
kicks. In the timescale of the experiment ($150$ kicks), the deviation
from horizontality must be less than $0.1^{\circ}$. This decoherence
effect is rather important. To the best of our knowledge, its importance
was not fully appreciated in previous experiments. A detailed discussion
of this effect will be presented elsewhere~\cite{gravity:tbp}.

\subsection{Conditions for the observation of localization effects}

We now discuss the conditions that must be satisfied in order to observe
localization effects experimentally.

Firstly, the system must present some kind of disorder: As discussed
in section \ref{sec:KRquasiper}, this means that $\kbar$, $\omega_{2}$
and $\omega_{3}$ and $\pi$ must be incommensurate. This is achieved
if we take $\kbar=2.89$, $\omega_{2}=2\pi\sqrt{5}$ and $\omega_{3}=2\pi\sqrt{13}$.
A more detailed discussion concerning the choice of these parameters
will be given in section \ref{sec:universality}.

\begin{figure}
\begin{centering}
\psfrag{x}{$t$ } \psfrag{y}{$\langle \pred^{2}\rangle$} \psfrag{a}{}
\includegraphics[width=7cm]{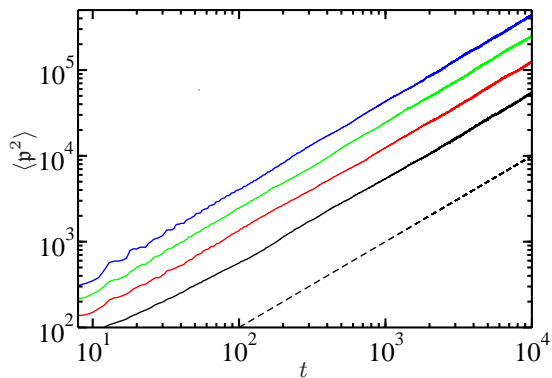} 
\par\end{centering}

\caption{\label{fig:simudiffchaoscl} Classical chaotic diffusion for the quasiperiodic
kicked rotor Eq.~(\ref{eqKR3DquasiperH}). The dynamics of an initial
thermal distribution of classical particles is simulated and the corresponding
mean kinetic energy is plotted versus time (number of kicks). The
stochasticity parameter $K$ (the modulation amplitude $\epsilon$)
varies linearly between $4$ and $9$ ($0.1$ and $0.8$), following
the experimental path, Fig.~\ref{fig:K-Epsilon}. The dashed
line of slope $1$ demonstrates the linear increase of $\langle \pred^{2}\rangle$
vs. time $t$. No classical localization effects are observed. The
chaotic diffusion is characteristic of the presence of pseudo-disorder
in the quasiperiodic kicked rotor, leading to a pseudo-random walk
in momentum space.}

\end{figure}

Secondly, in order to observe dynamical localization effects instead of
trivial classical localization, we must
be in a regime where the classical system has no KAM barriers which
can prevent the classical diffusive transport. For the standard, periodic
KR, full chaos is obtained for $K\gtrsim4$. In order to determine
the corresponding threshold for the quasiperiodic system, we performed
numerical simulations of the classical dynamics corresponding
to Eq.~(\ref{eqKR3DquasiperH}), for various values of the stochasticity
parameter. The dynamics is found to be fully diffusive
for $K\gtrsim2$, a considerably smaller value than for the standard
KR. In particular, no classical localization effects due to KAM barriers
are observed for $K\gtrsim2$. In any case, the experiments and the
numerical simulations in the following are all performed for $K>4,$
where the classical dynamics is diffusive, see Fig.~\ref{fig:simudiffchaoscl}.

Thirdly, short enough pulses must be used that they can be considered
as delta pulses \cite{Raizen:QSO96}. Numerical simulations of the
quasiperiodic kicked rotor with a finite pulse duration $\tau=0.9\mathrm{\mu s}$
and a thermal initial momentum distribution show that less than $1\%$
of the atoms are sensitive to the duration of the pulses. Only atoms
in the tails of the momentum distribution have sufficiently large
atomic velocity to move by a significant fraction of $\lambda_{L}$
during the pulse, thus feeling a smaller effective kick.

Fourthly, a sufficiently narrow initial state must be prepared in
order to observe dynamical localization, i.e. the freezing of the
initial diffusive expansion of the wave-function into an exponentially
localized state. A sufficient condition is that the initial width
of the momentum distribution be smaller than the localization length.
In our system, we have an initial momentum distribution of half-width
2$\kbar.$ This is comparable to the shortest localization
length at the lowest $K=4$ value, as experimentally proved, see inset 
of Fig.~\ref{fig:expeKineticEnergy}. 
A consequence is that, in this regime,
the exponential shape of the wavefunction after dynamical localization
is established is slightly rounded at the tip. For higher values -- say $K>5,$
-- the initial width of the atomic wavefunction can be safely neglected.  

Finally, decoherence processes must be kept small during 
the experiment. The large detuning of the standing wave allows
to keep the spontaneous emission rate very small, i.e. the corresponding
phase-coherence large as compared to the duration of the experiment.
A good control on the horizontality of the standing wave insures that
gravity do not lead to a destruction of localization effects on the
time-scale of the experiment.

\section{Experimental observation of the disorder induced metal-insulator
transition\label{sec:crossover}}

\begin{figure}
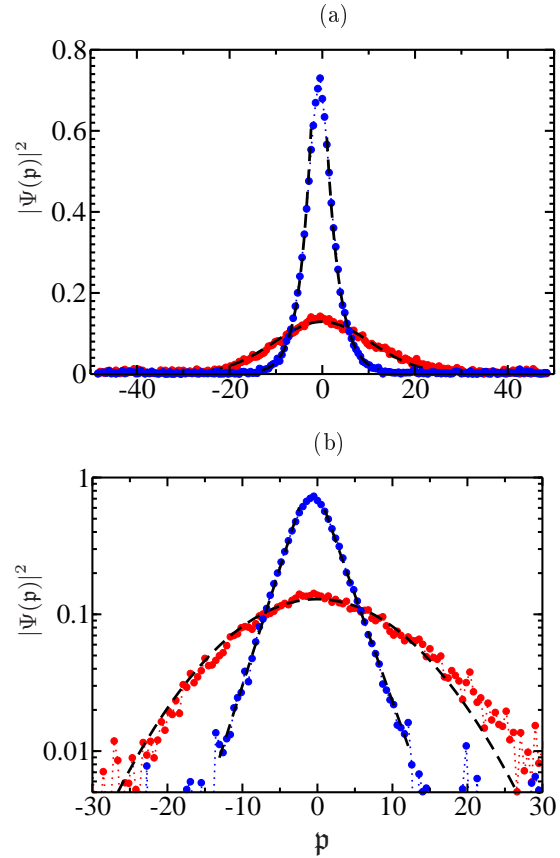

\begin{centering}
\psfrag{x}{} 
\psfrag{y}{$\vert\Psi(\pred)\vert^{2}$ } 
\psfrag{a}{(a)}
\psfrag{b}{(b)} 
\includegraphics[width=7cm]{figure04-1} 
\par\end{centering}

\begin{centering}
\psfrag{x}{{\large $\pred$ }}
\psfrag{y}{$\vert\Psi(\pred)\vert^{2}$ } 
\psfrag{a}{(a)}
\psfrag{b}{(b)} 
\includegraphics[width=7cm]{figure04-2} 
\par\end{centering}

\caption{\label{fig:expedistrib}(Color online) Experimentally measured
momentum distributions after 150 kicks, exponentially
localized in the insulator region (blue) and Gaussian in the diffusive
(metallic) region (red). (a) linear scale, (b) log scale. For both
curves $\kbar=2.89$, for the localized distribution (blue) $K=5.0$
and $\epsilon=0.24$, for the Gaussian distribution (red) $K=9.0$ and
$\epsilon=0.8$.}

\end{figure}

In a typical experimental run, we apply a sequence of kicks to the
atomic cloud and measure its dynamics. In the localized regime, the
evolution of its momentum distribution is ``frozen'' after the
localization time (typically of the order of 12 kicks at low $K$) 
into an exponential curve 
$\exp\left(-|\pred|/{\ell}\right)$.
In the diffusive regime, the initial Gaussian shape is preserved and
the distribution gets broader as kicks are applied, corresponding
to a linear increase of the average kinetic energy. Figure \ref{fig:expedistrib}
shows the experimentally observed momentum distributions, an exponentially
localized distribution for small $K$ and $\epsilon$ (blue curve),
characteristic of dynamical localization, and a broad, Gaussian-shaped
distribution for large $K$ and $\epsilon$ (red curve), characteristic
of the diffusive regime. %

Measuring the whole momentum distribution takes too much time: one must
repeat the whole sequence (from the preparation of a new atom cloud
up to the Raman measurement of the velocity distribution)
for each velocity class. Moreover, for each time step,
a complete momentum distribution must be measured. Fortunately,
it is sufficient, and much easier, to measure the population $\Pi_{0}(t)$
of the zero velocity class, as $\Pi_{0}^{-2}(t)$ is proportional
to $\langle \pred^{2}\rangle(t)$ (the total number of atoms is constant).
The proportionality factor between $\Pi_{0}^{-2}(t)$ and $\langle 
\pred^{2}\rangle(t)$
depends on the detailed shape of the momentum distribution and is
thus different in the localized and diffusive regime, but this small
difference is a small correction to the main phenomenon: divergence
of the localization length near the transition.

Note that, strictly speaking, the proportionality between $\Pi_{0}^{-2}(t)$
and $\langle \pred^{2}\rangle(t)$ breaks at criticality due to the multifractal
character of critical states \cite{Mirlin:RevModPhys08}. However,
on the time scale of the experiment ($t=150$ kicks), the deviation
from strict proportionality is seen (numerically) to be negligible.
At longer times (thousands or millions of kicks), the effect of multifractality
is visible and quantitatively measurable. This is beyond the scope
of this paper and will be analyzed elsewhere~\cite{multifractality:tbp}.

For each run, a value of $\Pi_{0}(t)$ is recorded after a given
number of kicks is applied, then the measurement sequence starts again
with the next number of kicks. We also record the background signal
obtained by not applying the Raman detection sequence, and the total
number of atoms in the cold-atom cloud. These signals are used to
correct the experimental data from background signals and long-term
drifts of the cloud population. %
\begin{figure}
\begin{centering}
\psfrag{t}{$t$} \psfrag{Pi}{$\Pi_{0}^{-2}$ } 
\includegraphics[width=7cm]{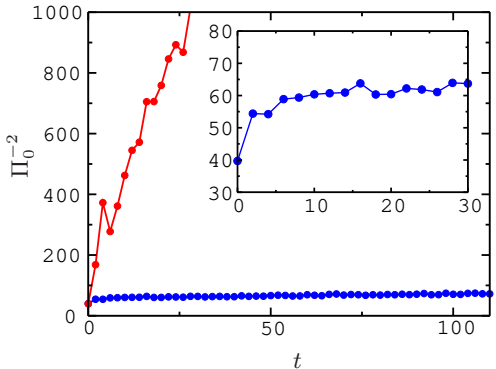} 
\par\end{centering}

\caption{\label{fig:expeKineticEnergy} (Color online) Temporal dynamics of
the quasi-periodic kicked rotor. We experimentally measure the
population $\Pi_{0}(t)$ of the zero-momentum class as a function
of time (number of kicks) and plot the quantity 
$\Pi_{0}^{-2}(t)\propto\langle \pred^{2}\rangle(t)$.
Clearly, it tends to a constant in the localized regime (blue lower
curve corresponding to $K=4$ and $\varepsilon=0.1$) and increases
linearly with time in the diffusive regime (red upper curve corresponding
to $K=9$ and $\varepsilon=0.8$). The inset shows the behavior close to
the localization time. $\kbar=2.89$.}

\end{figure}

Figure \ref{fig:expeKineticEnergy} shows the experimentally measured
$\Pi_{0}^{-2}(t)$ in the localized and diffusive regimes. It clearly
shows the initial diffusive phase and the freezing of the quantum
dynamics in the localized regime (blue curve in Fig.~\ref{fig:expeKineticEnergy}).
Along with the observation of an exponential localization of the wave-function
in Fig.~\ref{fig:expedistrib}, this constitutes a clear-cut proof
of the observation of dynamical localization. In the diffusive regime,
$\Pi_{0}^{-2}(t)$ is seen to increase linearly with time (red curve
in Fig.~\ref{fig:expeKineticEnergy}), corresponding to the Gaussian
red curve in Fig.~\ref{fig:expedistrib}.

\begin{figure}
\begin{centering}
\includegraphics[width=6cm]{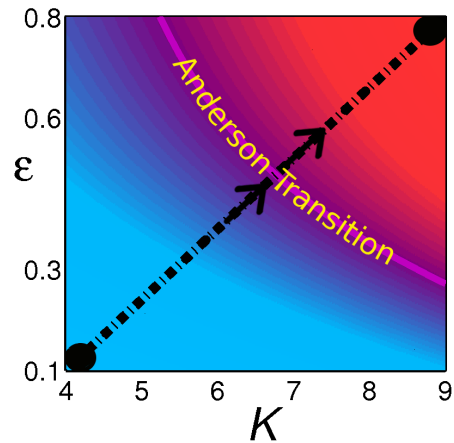} 
\par\end{centering}

\caption{\label{fig:K-Epsilon} (Color online) Phase diagram of the quasiperiodic
kicked rotor, from numerical simulations. The localized (insulator)
region is shown in blue, the diffusive (metallic) region is shown
in red. The experimental parameters are swept along the diagonal dash-dotted
line.}

\end{figure}

After having observed Anderson localization for strong effective disorder
strength and diffusive transport for small effective disorder, 
the next step is to walk the way between
these two regimes, and explore the phase transition expected (numerically)
to take place along a critical line in the plane $(K,\epsilon>0)$
(Fig.~\ref{fig:K-Epsilon}). In order to confine the transition
to a narrow range of parameters, we choose a path that
cross the critical curve (Fig.~\ref{fig:K-Epsilon}) {}``at a right
angle''; we thus vary simultaneously $K$ and $\varepsilon$ along
a line going from $K=4$, $\varepsilon=0.1$ in the localized region
to $K=9$, $\varepsilon=0.8$ in the diffusive region; the critical
line is then crossed at $K=K_{c}=6.6.$ %

A simple way to investigate the phase transition is the following
\cite{Shepelyansky:PRL89}. In the localized regime, wait for a time
longer than the localization time so that a localized frozen wave-function
is observed, then measure its localization length. One can in such
a way study the behavior of the localization length vs. disorder:
at criticality, it should diverge as ${\ell}\sim(K-K_{c})^{-\nu}$.
This would give the critical stochasticity parameter $K_{c}$ and
the critical exponent $\nu$. However, we cannot proceed that way
in our case, because when one approaches the critical point from the
insulator side, the localization time diverges as $\tau_{\ell}\sim\ell^{3}\sim(K-K_{c})^{-3\nu}$
in three dimensions (see below). In our system, a
localized momentum distribution would be observable in the vicinity
of the transition only for prohibitively large numbers of kicks, which
are, in practice, limited to $150$, essentially because of decoherence
effects and because the free fall of the atom cloud takes it out of
the standing wave. Consequently, it is vain to investigate experimentally
the Anderson transition only from \textit{static} properties such
as the divergence of the localization length at criticality, which
could be obtained only for $t\gg\tau_{\mathrm{loc}}$. Fortunately,
there is another way to observe the Anderson transition, which we
shall present in the following sections.

\section{Characterization of the Anderson phase-transition\label{sec:CharacterizationPhaseTrans}}

Finite-time effects act as finite-size effects do
on finite-size samples subjected to phase-transitions. Numerical
simulations of the Anderson transition on the standard 3D-Anderson
model are necessarily performed on finite-size samples of finite
size $L$. In the vicinity of the transition, the localization length
$\ell$ {[}see Eq.~(\ref{eq:Divergence}){]} diverges and thus can
greatly exceed $L$. In this regime, $L$ acts as an upper bound for
the effectively observed localization length $\ell_{L}$. This smooths
the transition, no divergence of the localization length can be 
\textit{directly}
observed on a finite-size sample. 
In order to overcome
this limitation, a powerful real-space renormalization method, called
finite-size scaling \cite{Pichard:Sarma:JPC81,MacKinnon:Kramer:PRL81},
was introduced. This method is based on a single parameter scaling
hypothesis \cite{Abrahams:PRL79} and allows to extrapolate from the
scaling behavior of $\ell_{L}$ versus $L$ the asymptotic
value of the localization length $\ell$ corresponding to $L\rightarrow\infty$.
We can generalize static scaling laws to cover our time-dependent
problem (see \cite{Stauffer:94} for a similar approach in percolation
theory). The single parameter scaling theory \cite{Abrahams:PRL79},
successfully used for the standard 3D Anderson model \cite{Pichard:Sarma:JPC81,MacKinnon:Kramer:PRL81},
can be applied to analyze our experimental and numerical data, and
especially to determine the critical properties of the Anderson transition
that we observe, i.e. the critical exponents.

\subsection{Scaling law at finite time}

\label{sec:Scalinglaw}
Knowing the asymptotic behavior when $t\rightarrow\infty$
is not enough, an additional time-dependent property is needed, too,
which we shall investigate now. For $K$ far above $K_{c}$ one observes
normal diffusion, $\langle p^{2}\rangle\propto t$, whereas for $K$
far below $K_{c}$, the quantum dynamics freezes, at sufficiently
long times. Following the standard analysis of the Anderson transition,
we make the hypothesis that the transition that we observe for the
quasi-periodically kicked rotor follows a one-parameter scaling law~\cite{Ohtsuki:AndersonTrans:JPSJ97} 
(the validity of this scaling hypothesis will of course be checked 
at the end of the analysis). 
At the critical point, a third
kind of dynamics, namely anomalous diffusion, 
with $\langle p^{2}\rangle\sim t^{k}$
$k\neq1$, is expected. Let us consider in greater detail the behavior
very close to $K_{c}$ where these three different laws merge.

In the localized regime, for sufficiently long times, the behavior
depends only on the localization length which diverges as $K$ goes
to $K_{c}$: 
\begin{equation}
\langle \pred^{2}\rangle\sim{\ell}^{2}\sim(K_{c}-K)^{-2\nu}\;
\ \ \ \ \ \ \ (\mathrm{for}\ K<K_{c})\;,
\label{eqnu}
\end{equation}
with $\nu$ the localization length critical exponent. 

For $K>K_{c}$,
the mean kinetic energy increases linearly with time, and the proportionality
constant is the diffusion coefficient $D(K)$. For $K<K_{c}$, 
$\langle \pred^{2}\rangle$
is bounded by Eq.~(\ref{eqnu}) and there is no diffusion. 
Thus $D(K)$ vanishes below $K_{c}.$ A different critical exponent $s$ is used
to describe how $D(K)$ goes to zero above threshold: 
\begin{equation}
D(K)\sim(K-K_{c})^{s}\;\ \ \ \ \ \ \ (\mathrm{for}\ K>K_{c}).\label{eqs}
\end{equation}

We shall now find a single expression presenting these
two limit behaviors and also displaying anomalous diffusion at the
critical point. We note that, according to the theory of phase-transitions
in finite-size samples, a scaling can be applied to 
$\langle \pred^{2}\rangle$ depending on the two variables $1/t$ and $(K-K_{c})$, 
both going to zero. 
We thus use the general scaling law:
\begin{equation}
\langle \pred^{2}\rangle=t^{k_{1}}F\left[\left(K-K_{c}\right)t^{k_{2}}\right]\;,
\label{eqscalinglaw}
\end{equation}
with $F(x)$ an unknown scaling function. The exponents
$k_{1}$ and $k_{2}$ can be determined as follows.

In the diffusive regime, for long enough times, we
must recover the diffusion law with
$D\sim(K-K_{c})^{s}$ {[}Eq.(\ref{eqs}){]};
hence, for $x\gg1$, the scaling function $F(x)$
should scale as $x^{s}$: 
\begin{equation}
\langle \pred^{2}\rangle\sim t^{k_{1}+sk_{2}}\left(K-K_{c}\right)^{s}\;.
\label{eqasymptote}
\end{equation}
As in the diffusive regime, $\langle \pred^{2}\rangle\sim t$,
we must have $k_{1}+sk_{2}=1$.

In the localized regime, on the other hand, one must
recover $\langle \pred^{2}\rangle\sim(K_{c}-K)^{-2\nu}$ {[}Eq. (\ref{eqnu}){]}
for sufficiently long times. 
Thus, for $x\rightarrow-\infty$,
$F(x)\rightarrow(-x)^{-2\nu}$, giving: 
\begin{equation}
\langle \pred^{2}\rangle=t^{k_{1}-2\nu k_{2}}(K_{c}-K)^{-2\nu}
\end{equation}
which is compatible with Eq.~(\ref{eqnu}) only if $k_{1}=2\nu k_{2}.$
These two relations determine $k_{1}$ and $k_{2}$ in terms of the
physically more meaningful critical exponents $s$ and $\nu$.

\begin{eqnarray*}
k_{1} & = & \frac{2\nu}{s+2\nu}\\
k_{2} & = & \frac{1}{s+2\nu}.\end{eqnarray*}
In the standard Anderson model, the critical exponents are related
by Wegner's scaling law \cite{Wegner:ZFP76}: 
\begin{equation}
s=(d-2)\nu\;,\label{eqwegnerscalinglaw}
\end{equation}
with $d$ being the dimensionality of the system. For our system,
one obtains:
\begin{equation} 
k_{1}=2/3;\ \ \ \ \ \ \ \ k_{2}=1/3\nu.
\end{equation} 
We therefore expect
at the critical point anomalous diffusion 
with $\langle \pred^{2}\rangle=t^{k_{1}}F(0)\sim t^{2/3}$.
We present in the next sub-section a numerical and experimental validation
of this prediction.

\subsection{Critical anomalous diffusion}

\begin{figure}
\begin{centering}
\psfrag{x}{$t$} \psfrag{y}{$\langle \pred^{2}\rangle$} 
\includegraphics[width=0.9\linewidth]{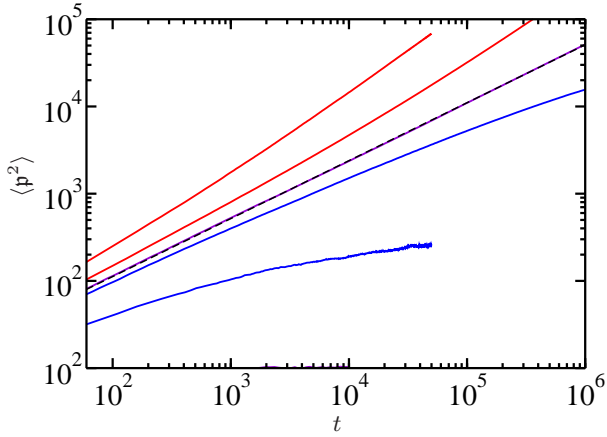} 
\end{centering}

\caption{\label{fig:anno-diff-simu} 
(Color online) Numerically simulated time-evolution
of $\langle \pred^{2}\rangle$ for the quasiperiodic kicked rotor. At
the critical point $K=K_{c}\approx6.4$ (purple middle curve), anomalous
diffusion $\langle \pred^{2}\rangle\sim t^{2/3}$ is clearly observed,
as expected from theoretical arguments (cf. text). The log-log plot
of the critical curve is very well fitted by a straight line of slope
$0.664$ (black dashed line). In the vicinity of the transition, the
dynamics departs from the anomalous diffusion to tend gradually either
to a diffusive dynamics (red upper curves corresponding 
to $K=> K_{c}$
bending upwards for large $t$) or to a localized dynamics
(blue lower curves corresponding to $K< K_{c}$ bending
downwards for large $t$). Other
parameters are $\kbar=2.85$, $\omega_{2}=2\pi\sqrt{5}$
and $\omega_{3}=2\pi\sqrt{13}$.}

\end{figure}

We verified numerically that the critical behavior, corresponding
to the anomalous diffusion in $t^{2/3}$ is observed up to a very
large number of kicks ($t=10^{6}$). 
The (purple) middle curve of Fig.~\ref{fig:anno-diff-simu}
displays the time-evolution of $\langle \pred^{2}\rangle$ from numerical
simulations for the stochasticity parameter $K=6.4$. Anomalous diffusion
$\langle \pred^{2}\rangle\sim t^{2/3}$ is clearly seen from the log-log
plot over 4 orders of magnitude, which is very well fitted by a straight
line of slope $0.664$. Other curves, for different $K$, tend at
long times to bend either horizontally (below $K_{c}$) or towards
slope unity (above $K_{c}$). This is a clear proof
that we face here a true phase transition and not a smooth cross-over. 
Note also that
the fact that the numerically measured critical slope is very close
to the theoretical prediction 2/3 implies that the Wegner's scaling
law $s=\nu$ is valid at an accuracy better than 1\%.

\begin{figure}
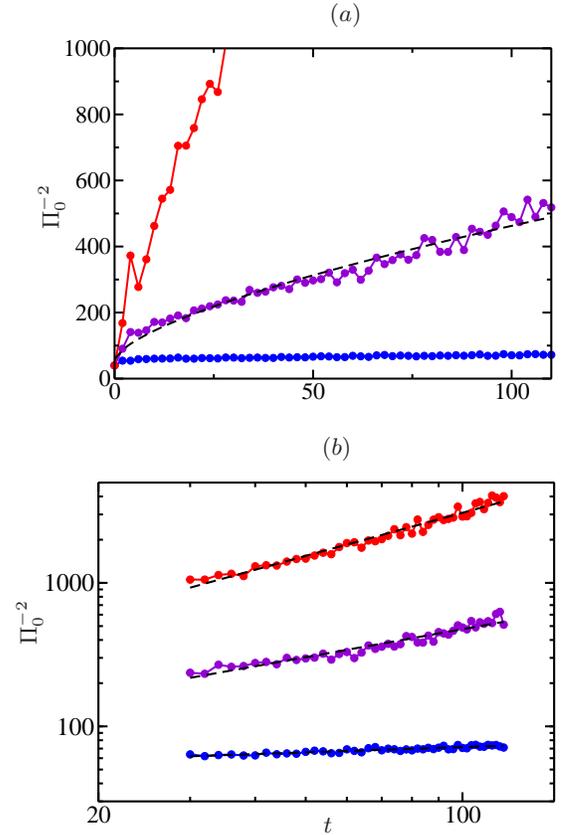

\begin{centering}
\psfrag{a}{$(a)$} \psfrag{t}{} \psfrag{Pi}{$\Pi_{0}^{-2}$ }
\includegraphics[width=7cm]{figure08-1} 
\par\end{centering}

\begin{centering}
\psfrag{b}{$(b)$}\psfrag{x}{$t$} \psfrag{y}{$\Pi_{0}^{-2}$}
\includegraphics[width=7cm]{figure08-2} 
\par\end{centering}

\caption{\label{fig:expeanodiff} (Color online) 
Experimentally observed time-evolution
of $\Pi_{0}^{-2}\sim\left\langle \pred^{2}\right\rangle $ 
for the quasiperiodic
kicked rotor. Close to the critical point $K=K_{c}\approx6.4$ (purple
middle curve), anomalous diffusion $\Pi_{0}^{-2}(t)\sim t^{2/3}$
is clearly observed. (a) The critical anomalous curve is well fitted
by $\Pi_{0}^{-2}(t)=A+Bt^{2/3}$ (black dashed line). The red upper
curve evidence the far-above-criticality diffusive behavior ($K=9.0$)
and the blue lower curve the far-below-criticality ($K=4.0$) localized
behavior. (b) These experimental results show a clear algebraic behavior,
with exponent $\approx0$ (blue lower curve, localized regime), 2/3
(purple middle curve, critical regime) and 1 (red upper curve, diffusive
regime), slightly perturbed by decoherence processes responsible for
the residual increase in the localized regime. Other
parameters are the same as in Fig.~\ref{fig:anno-diff-simu}.}

\end{figure}

Fig.~\ref{fig:expeanodiff} displays the experimental evolution of
$\Pi_{0}^{-2}(t)\sim\langle \pred^{2}\rangle$ versus time. The critical
curve (middle curve corresponding to $K\approx6.4$) in purple is
well fitted by the relation $\Pi_{0}^{-2}(t)=A+Bt^{2/3}$, see Fig.~\ref{fig:expeanodiff}a.
Fig.~\ref{fig:expeanodiff}b displays in log-log scale the experimental
data $\Pi_{0}^{-2}(t)$ vs $t$. The algebraic dependence (with exponent
$2/3$) of the critical dynamics is again clearly visible. In all
plots in Figs. \ref{fig:expeanodiff} the red upper curves evidence
the above-criticality diffusive behavior and the blue lower curves
the below-criticality localized behavior.

From renormalization theory, we know that the critical behavior 
shows the existence of a fixed hyperbolic point~\cite{Kramer:Localization:RPP93}.
It is a fixed point because the critical behavior remains the same
at all times (opposite to the localized case for example, for which
a characteristic time can be defined, the localization time), and
it is a hyperbolic point since the localized dynamics close
to criticality will follow only for a finite time the anomalous diffusion
with exponent $2/3$ and will progressively tend to a localized behavior
for large enough time. The rate at which the behavior changes is related
to the critical exponent of the phase transition $\nu$.

\begin{figure}
\begin{centering}
\psfrag{x}{$t$ } 
\psfrag{y}{$\ln\Lambda$ } 
\includegraphics[width=7cm]{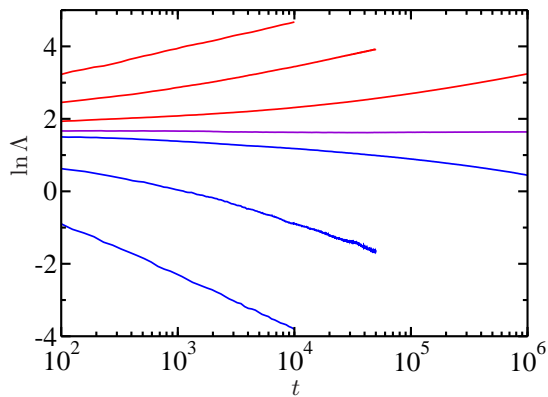} 
\end{centering}

\caption{ 
\label{fig:simulnlambdavst} 
(Color online) Numerical
simulation showing the evolution of the dynamics
from the critical behavior towards either a diffusive
dynamics or a localized state. 
Plotting the quantity $\ln\Lambda=\ln (\langle \pred^{2}\rangle t^{-2/3})$
vs. $\ln t$ allows to easily distinguish the critical behavior from
diffusive or localized behavior: The critical curve (corresponding
to $K=K_{C}\approx6.4$) has a zero slope; whereas the far localized
($K=4.0$) one has a slope $-2/3$ and the far diffusive ($K=9.0$) one
a slope $1/3$. Other parameters are the same as in
Fig.~\ref{fig:anno-diff-simu}.
}
\end{figure}

An efficient way to observe the departing of the dynamics from the
critical anomalous diffusion is to consider the quantity 
\begin{equation}
\Lambda=\frac{\langle \pred^{2}\rangle}{t^{2/3}}\;,\label{eqLambda}
\end{equation}
 or, equivalently, in the case of the experimental data: 
 \begin{equation}
\Lambda_{0}=\frac{1}{\Pi_{0}^{2}(t)t^{2/3}}\;,\label{eqLambdaPi0}
\end{equation}
as a function of time. This is illustrated in Fig.~\ref{fig:simulnlambdavst},
which displays $\ln\Lambda$ vs. $t$. The critical behavior can be
easily pin-pointed: The corresponding (purple) curve has a zero slope,
as the quantity $\Lambda$ is (asymptotically) constant at criticality.
In the diffusive regime, the quantity $\Lambda$ increases with time
(red curves), whereas it decreases in the localized regime (blue
curves). In the far localized regime, we observe an algebraic dependence
$\Lambda(t)\sim t^{-2/3}$ as $\langle \pred^{2}\rangle(t)=2{\ell}^{2}$
for $t>\tau_{\mathrm{loc}}$. In the far diffusive regime, the algebraic
dependence is $\Lambda(t)\sim t^{1/3}$ as $\langle \pred^{2}\rangle(t)\sim t$.

The above numerical and experimental observations validate the theoretical
prediction for the critical behavior: $\langle \pred^{2}\rangle\sim t^{2/3}$.
Such critical behavior for the quasi-periodic kicked rotor was predicted
using another scaling approach and numerically verified in \cite{Shepelyansky:PD97}.
It was also numerically observed for a spatially-3D kicked 
rotor \cite{GarciaGarcia:PRE09}
and in the standard 3D Anderson model \cite{Ohtsuki:AndersonTrans:JPSJ97},
and put on
firm theoretical grounds in \cite{Fyodorov:Andersoncritical:PRL94}.

\subsection{Finite-time scaling\label{sub:FTS}}

We shall now explain the procedure
used to verify the scaling of our numerical and experimental
data according to the law deduced above: 
\begin{equation}
\langle \pred^{2}\rangle=t^{2/3}F\left[(K-K_{c})t^{1/3\nu}\right]\;.
\label{eq:scalinglawassumption}
\end{equation}
Our method is similar to the finite-size scaling procedure used by
MacKinnon and Kramer \cite{MacKinnon:Kramer:PRL81,Kramer:FSS:ZPB83},
and Pichard and Sarma \cite{Pichard:Sarma:JPC81} to numerically study
the Anderson transition on finite-samples of the 3D Anderson model,
but we apply it here to the temporal behavior of the
data, thus the name ``finite-\emph{time} scaling''.

We assume the quantity $\Lambda(K,t)=\langle p^{2}\rangle t^{-2/3}$
to be an arbitrary function 
\begin{equation}
\Lambda(K,t)=f\left(\xi(K)t^{-1/3}\right)\;,\label{eqscaling}
\end{equation}
where the scaling parameter $\xi(K)$ depends \textit{only} on $K$,
which is the parameter appearing in the one-parameter
scaling hypothesis. This scaling assumption is less restrictive
than Eq.~(\ref{eq:scalinglawassumption}) since no assumption on
the dependence of $\xi$ on $K$ is made. We must thus
show that the resulting scaling parameter $\xi(K)$ is compatible
with Eq.~(\ref{eq:scalinglawassumption}).

\begin{figure}
\begin{centering}
\includegraphics[width=0.9\linewidth]{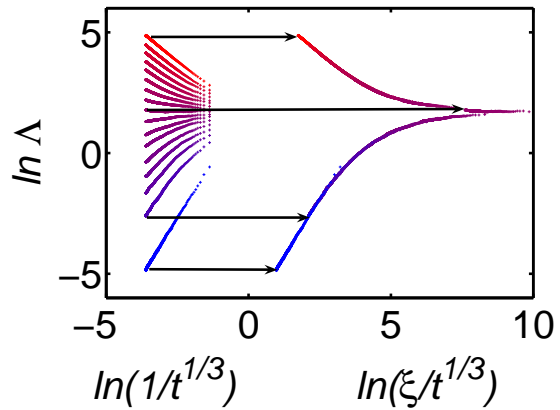} 
\par\end{centering}

\caption{\label{fig:principefinitetimescaling} 
(Color online) Raw
numerical data, displayed in the form 
$\ln\Lambda=\ln\langle(\pred^{2}\rangle t^{-2/3})$
vs. $\ln t^{-1/3}$ (on the left). Each curve corresponds to a different
stochasticity parameter $K$. The finite-time scaling procedure consists
in shifting horizontally each curve by a quantity $\ln\xi(K)$ so
that the curves overlap. This allows one to determine both the scaling
function $f$ (on the right) and the scaling parameter $\xi(K)$.}

\end{figure}

In the left part of Fig.~\ref{fig:principefinitetimescaling}
we display plots of $\ln\Lambda(K,t)$ vs. $\ln t^{-1/3}$ for different
values of $K$. For most values of $\ln\Lambda$,
several values of $\ln t^{-1/3}$ correspond to the same $K$ value.
The only way to conform with the condition (\ref{eqscaling})
is to shift each curve horizontally by a different quantity $\ln\xi(K)$
such that curves corresponding to different values
of $K$ overlap. This can be achieved by minimizing the variance
of the values $\ln\xi(K)t^{-1/3}$ corresponding to each value of
$\ln\Lambda$. The function $\xi(K)$ can be determined by applying
a least square fit to the data. 

This minimization procedure does not allow one to compute the absolute
scale of $\xi(K),$ as the shifting procedure (see Fig.~\ref{fig:principefinitetimescaling})
is invariant under a global shift of the origin. We can thus set
the scaling parameter $\xi(K)$ to be equal to the localization length
in the strongly localized regime where
the duration of the experiment is much larger than the localization
time, and $\langle \pred^{2}\rangle$ converges to its asymptotic
value $2{\ell}^{2}$.
Thus \begin{eqnarray*}
\Lambda(K,t) & = & f\left(\xi(K)t^{-1/3}\right)=2{\ell}^{2}t^{-2/3}\;,\end{eqnarray*}
which implies, if we identify the scaling parameter
with the localization length, $\xi(K)\sim\ell$, \[
f(x)=2x^{2}\;.\]

\begin{figure}
\begin{centering}
\includegraphics[width=7cm]{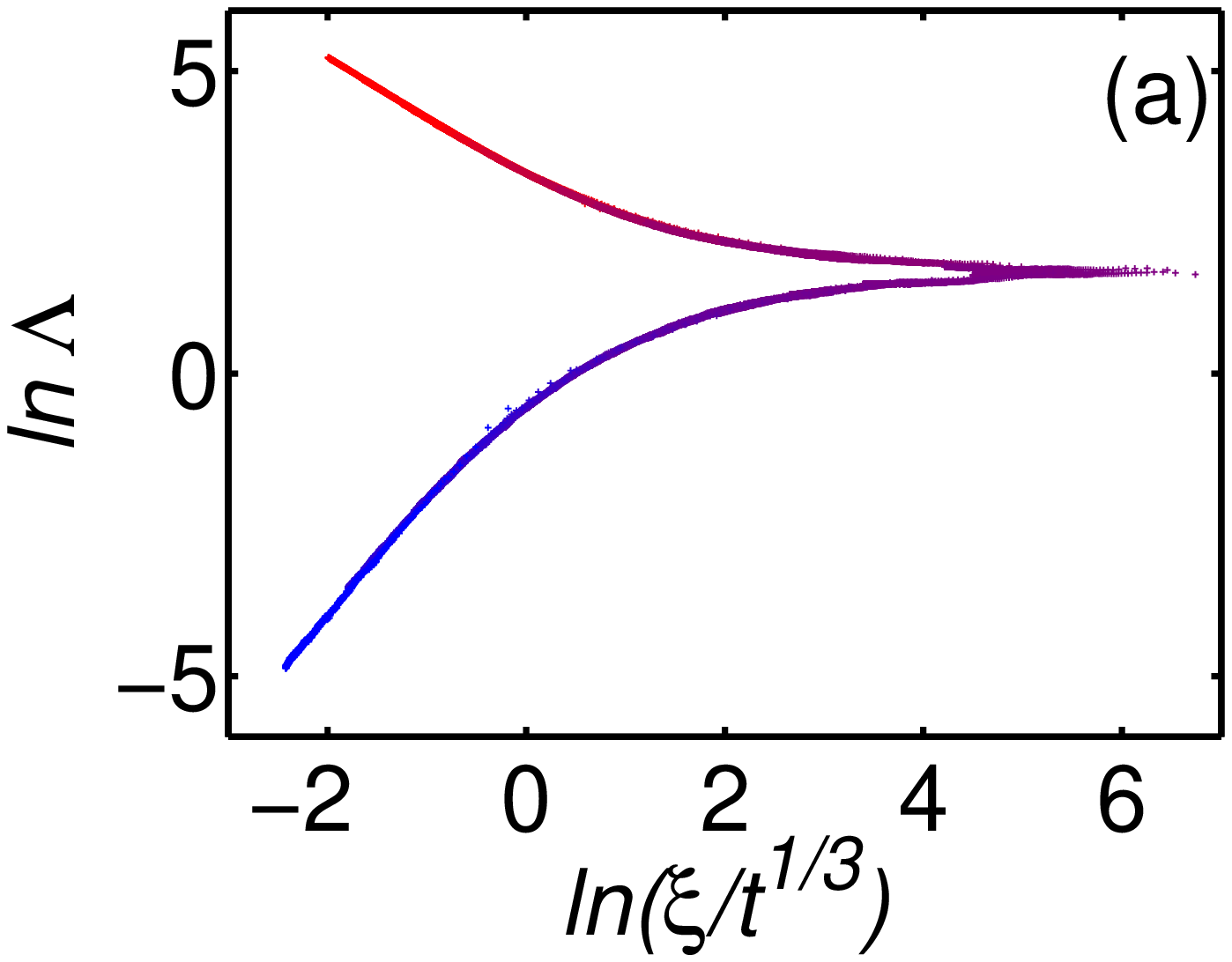}
\includegraphics[width=7cm]{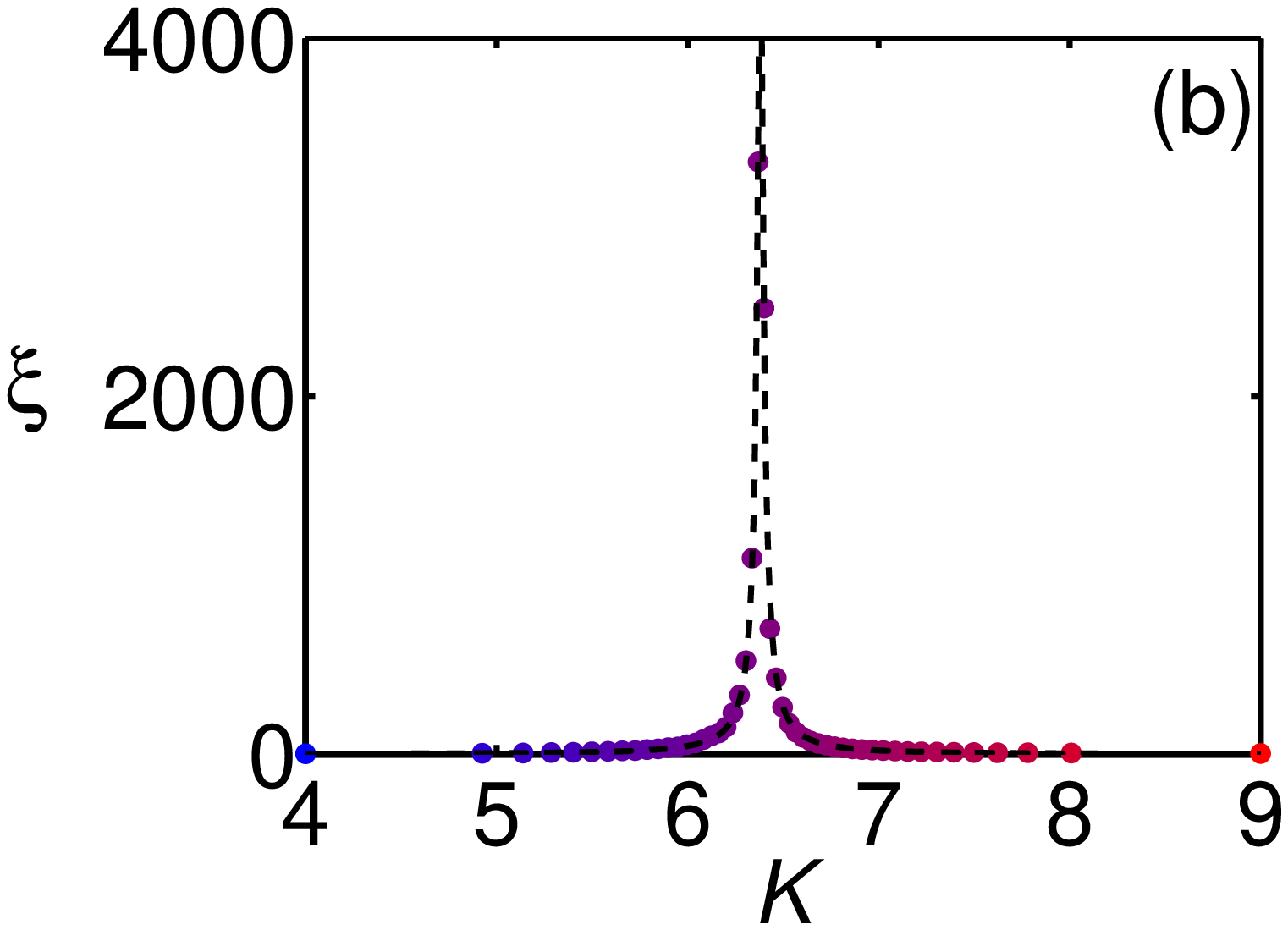} 
\par\end{centering}

\caption{\label{fig:simufts}(Color online) 
Finite-time scaling applied to
the results of numerical simulations of the quasiperiodic kicked rotor.
The time-evolution of $\langle p^{2}\rangle$ is computed as a function
of time, from $30$ to $10^{4}$ kicks, for several values of $K$
between $K=4$ and $K=9$. The finite-time scaling procedure allows
us to determine both the scaling function $f$ (a), clearly
displaying an upper branch (red) associated with the diffusive regime,
and a lower branch (blue) associated with the localized regime. The
dependence of the scaling parameter $\xi$ on $K$ (b) displays a
divergent behavior around the critical point $K_{c}=6.4$, which is
the signature of the Anderson phase transition. The dashed line is
a fit using Eq.~(\ref{eq:xivsK}). The resulting critical exponent
is $\nu=1.6\pm0.1$. Other parameters are $\kbar=2.85$,
$\omega_{2}=2\pi\sqrt{5}$ and $\omega_{3}=2\pi\sqrt{13}$.}

\end{figure}

\begin{figure}
\begin{centering}
\includegraphics[width=7cm]{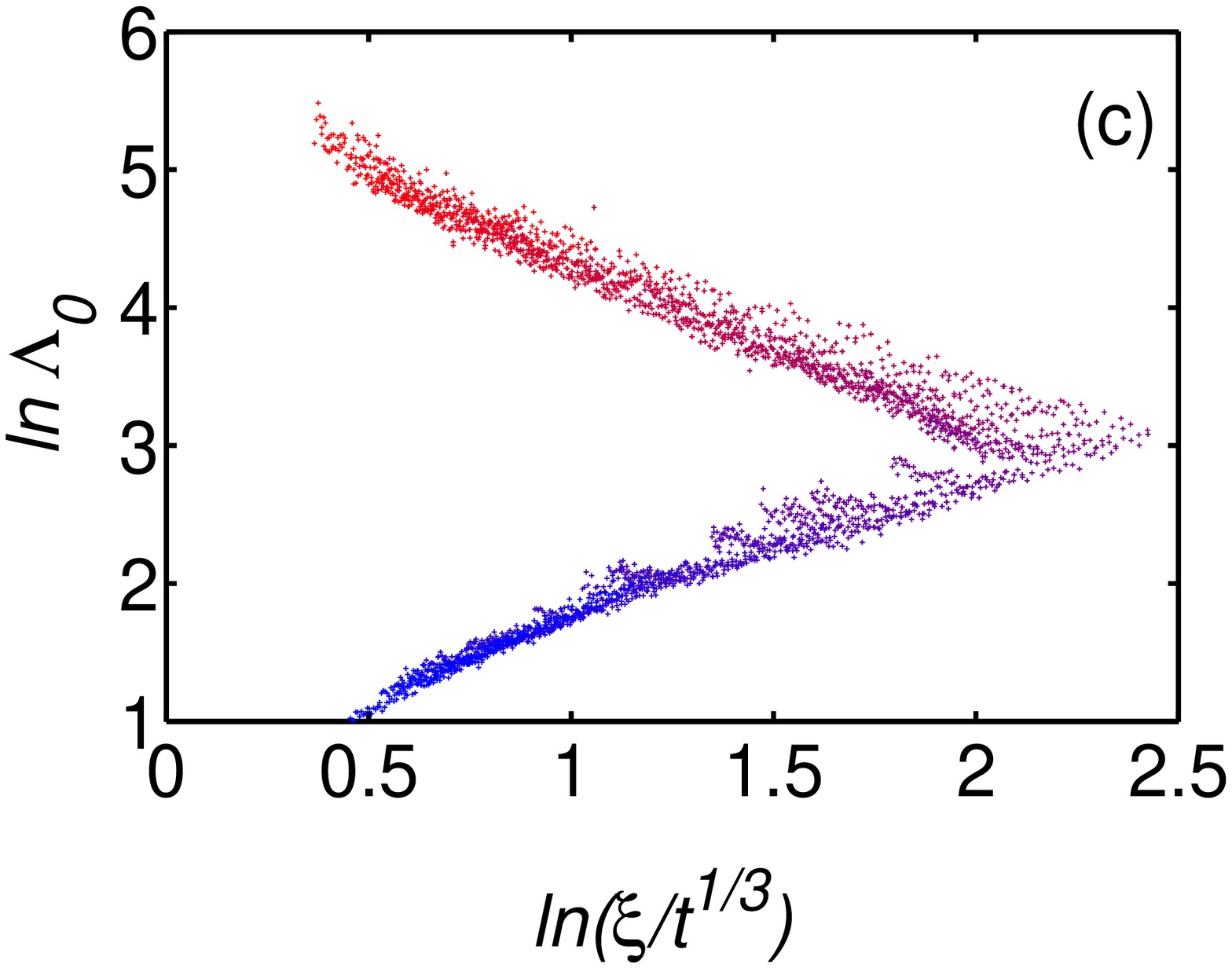} \includegraphics[width=7cm]{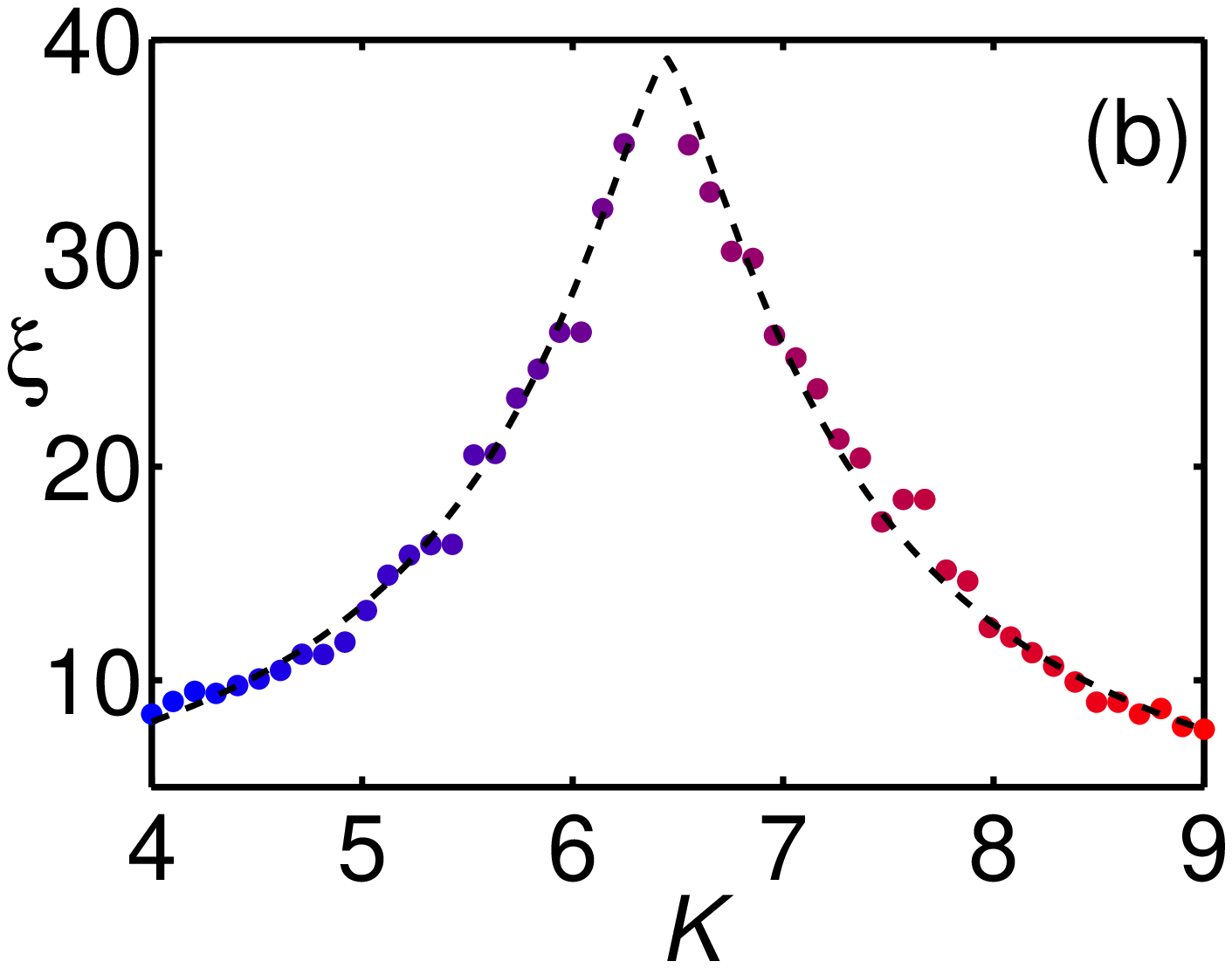} 
\par\end{centering}

\caption{\label{fig:expefts}(Color online) Finite-time scaling applied to
the experimental results (from $30$ to $150$ kicks). The scaling
procedure is the same as in Fig.~\ref{fig:simufts}. (a) The fact
that all experimental points lie on a single curve, with a diffusive
(red) and a localized (blue) branch, is a proof of the relevance of
the one-parameter scaling hypothesis. (b) The maximum displayed by
the scaling parameter $\xi$ in the vicinity of $K_{c}=6.4$ is a
clear-cut proof of the Anderson transition. Phase-breaking mechanisms
(cf. text) smooth the divergence at the critical point. When these
effects are properly taken into account, one obtains a critical exponent
$\nu=1.4\pm0.3$, {[}the dashed line is a fit with Eq.~(\ref{eq:xivsK}){]}
compatible with the numerical results. This plot corresponds to $48$
experimental runs.Other parameters as in Fig.~\ref{fig:simufts}.}

\end{figure}

Figures \ref{fig:simufts}(a) and \ref{fig:expefts}(a) show the results
of the fitting procedure applied to the numerical data and to the
experimental data, respectively. In both cases, the procedure groups
all points in a \emph{ single} curve, within the accuracy of the data.
The resulting curve clearly displays two branches, a diffusive (red)
and a localized (blue) one, with the critical point being at the tip
joining the two branches; this is a signature of the Anderson transition. 
It also justifies {\textit a posteriori} 
the scaling hypothesis Eq.~(\ref{eqscaling}) used for analyzing the data.

The scaling parameter $\xi(K)$ is plotted in Figs. \ref{fig:simufts}(b)
and \ref{fig:expefts}(b), for numerical and experimental data respectively.
As stated above, this parameter can be identified to the localization
length in the localized regime. In the diffusive regime, it scales
as the inverse of the diffusive constant. Indeed, in the far diffusive
regime one has $\langle \pred^{2}\rangle=D(K)t$, which implies \begin{eqnarray*}
\Lambda(K,t) & = & D(K)t^{1/3}\\
f(x) & = & x^{-1}.\end{eqnarray*}
 so that $\xi(K)=1/D(K)$ in the far diffusive regime.

One notes that $\xi(K)$ increases rapidly in the vicinity of the
critical value $K_{c}$, on both sides of the transition. This corresponds
to a divergence of the localization length and to a vanishing of the
diffusion constant at criticality (in practice smoothed
by decoherence, see below). This constitutes a clear {\emph{experimental}
evidence of the Anderson phase transition.

\subsection{Experimental determination of the critical exponent}

The behavior of $\xi(K)$ gives a fundamental information about the
transition, namely the value of the localization length critical exponent
$\nu$. There is a discrepancy in the literature between the theoretical
predictions $\nu=1$ \cite{Vollhardt:Wolfle:PRL82}, $\nu=1.5$ \cite{GarciaGarcia:PRL08},
and the result of numerical simulations $\nu=1.57\pm0.02$ \cite{Slevin:PRL99},
which stresses even more the importance of an experimental determination.
In this section, we present the first unambiguous experimental determination
of the critical exponent of the Anderson transition in 3 dimensions.

The finite-time scaling procedure allows us to extract
from finite-time experimental data the localization length ${\ell}$
(corresponding to $t\rightarrow\infty$), which is the order parameter
of the Anderson transition. It is given by the scaling parameter $\xi(K)$
and predicted to diverge at criticality with the power law 
\begin{equation}
{\ell}\sim\vert K-K_{c}\vert^{-\nu}\;.\label{eq:nu}
\end{equation}
 We thus expect that the singularity in $\xi(K)$ can be described
by Eq.~(\ref{eq:nu}), and to be able to extract the value of the critical exponent $\nu.$
This is of primary importance, as there is presently no unambiguous accurate 
experimental determination of $\nu$ for non-interacting particles, and 
there is a discrepancy in the literature between the theoretical
predictions $\nu=1$ \cite{Vollhardt:Wolfle:PRL82}, $\nu=1.5$ \cite{GarciaGarcia:PRL08},
and the result of numerical simulations $\nu=1.57\pm0.02$ \cite{Slevin:PRL99}.

When the slope of $\ln\Lambda$ vs $\ln t^{-1/3}$ is small,
as it is near the critical point, the scaling procedure tends to round
off the singularity in $\xi(K)$. Moreover, decoherence
in the experiment produces a cut-off the algebraic divergence. If
the system has a finite phase-coherence time $\tau_{\varphi}$, a
new characteristic length  \cite{Shapiro:scalingand:PRB81} $p_{\varphi}=\left[D\tau_{\varphi}\right]^{1/2}$ 
appears in the problem, which
sets an upper bound for the observable
localization length $\ell$ and thus smooths its divergence at criticality.
In practice, we model such smoothing by introducing a small cut-off
on the divergence of $\xi(K)$, which takes into account
both the finite-time scaling procedure itself and decoherence effects:
\begin{equation}
\frac{1}{\xi(K)}=\alpha\vert K-K_{c}\vert^{\nu}+\beta\;.\label{eq:xivsK}
\end{equation}

The experimental data have been fitted with this formula (\ref{eq:xivsK})
{[}dashed curve in Fig.~\ref{fig:expefts}(b){]}, 
which gives $K_{c}\simeq6.4\pm0.2$,
and a critical exponent $\nu=1.4\pm0.3$. In order to compare these
results to the ideal case of the perfectly coherent quasiperiodic
kicked rotor, Eq.~(\ref{eq:KRquasiper}), we also
fitted the curve in Fig.~\ref{fig:simufts}b with Eq. (\ref{eq:xivsK});
in this case, the cutoff $\beta$ accounts for limitations of the
finite-time scaling procedure. The model Eq.~(\ref{eq:xivsK}) fits
very well to the numerical data {[}dashed curve in Fig.~\ref{fig:simufts}b{]}
and gives the critical stochasticity $K_{c}\simeq6.4\pm0.1$ and the
critical exponent $\nu=1.6\pm0.2$. The good agreement between the
numerical simulations and the experimental results proves that spurious
effects (such as decoherence) are well under control. Moreover, the
experimental value we obtained $\nu=1.4\pm0.3$ is compatible with
the value found in numerical simulations of the true random 3D Anderson
model \cite{MacKinnon:JPC94,Slevin:PRL99}. We emphasize that there
are no adjustable parameters in our procedure, all parameters are
determined using the atoms themselves as probes.

\section{Universality of the Anderson transition \label{sec:nuetuniversality} }

At this point, a reasonable question is: Does the
quasiperiodic kicked rotor exhibits the same critical phenomena --
i.e. belongs to the same (orthogonal) universality class \cite{Mirlin:RevModPhys08}
-- as the true 3D-Anderson model. Can this simple three-frequency
dynamical system exactly mimic the critical behavior of 3D disordered
electronic conductors? In this section, 
we show  that the answer is positive: \textit{The
3-frequency quasiperiodic kicked rotor and the true 3D-Anderson model
belong to the same universality class}. This is a strong claim that
relies on a very precise determination of the critical exponent $\nu$. 
The accuracy of this
determination is comparable to that obtained in the most sophisticated
numerical studies of the 3D Anderson model \cite{Slevin:PRL99,Schreiber:EPJ00}.
Within numerical uncertainties, the critical exponent is found to
be universal and identical to the one found for the 3D-Anderson model
\cite{Slevin:PRL99}. The technical details of the calculation have
already been reported in \cite{Lemarie:UnivAnd:arxiv09}. We here just
discuss the essential ingredients proving universality.

\subsection{Precise estimate of the critical exponent}

Reliably distinguishing the different universality classes of the
Anderson transition requires a very precise determination
of the critical exponent; for instance, the value $\nu=1.43\pm0.04$
for the unitary symmetry class is close to the one for the orthogonal
symmetry class \cite{Slevin:PRL97} $\nu=1.57\pm0.02$. 

The main uncertainty in our experimental determination
of the critical exponent is due to statistical errors on $\Pi_{0}$
and to the limited duration of the experiment. However, numerical
simulations are not limited to 150 kicks but can be ran for several
thousands of kicks, and statistical uncertainties on $\langle \pred^{2}\rangle$
can be sharply reduced by averaging over initial conditions. The numerical
inaccuracy in the finite-time scaling determination of $\nu$ from
the numerical data is thus mainly due to the procedure failing to
reproduce the singular behavior of the scaling function at the critical
point.

How can one improve the accuracy on the determination
of the critical exponent $\nu$? This can be achieved by fitting
directly the raw data $\ln\Lambda(K,t)$. The starting point of our
analysis is the behavior of the scaling function $\mathcal{F}\equiv\ln F$
in the vicinity of the critical point: 
\begin{eqnarray}
\ln\Lambda & = & \ln F\left[\left(K-K_{c}\right)t^{1/3\nu}\right]\nonumber \\
 & = & \mathcal{F}\left[\left(K-K_{c}\right)t^{1/3\nu}\right]\;.
 \label{eq:scalinglnlambda}
 \end{eqnarray}
As $\ln\Lambda(K,t)$ is an analytical function for \emph{finite}
$t$ (Fig.~\ref{fig:simufts}), the scaling function $\mathcal{F}$
can be expanded around $K_{c}$: \begin{equation}
\ln\Lambda(t)\simeq\ln\Lambda_{c}+\left(K-K_{c}\right)t^{1/3\nu}\mathcal{F}_{1}+...\;,\label{eqlinscaling}\end{equation}
 where $\ln\Lambda_{c}\equiv\mathcal{F}[0]$ and
 $\mathcal{F}_{1}=\mathrm{d}\mathcal{F}(x)/\mathrm{d}x\vert_{x=0}$.

\begin{figure}
\begin{centering}
\psfrag{x}{$K$ } 
\psfrag{y}{$\ln\Lambda$ } 
\includegraphics[width=7cm]{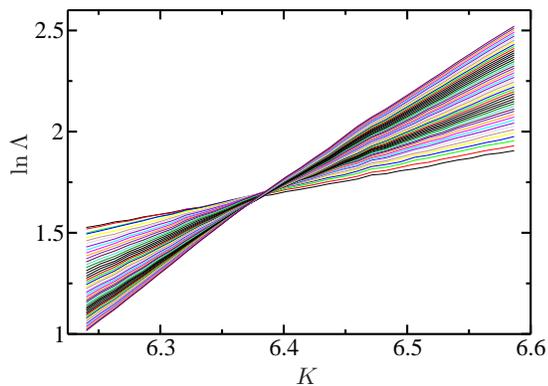} 
\end{centering}

\caption{\label{fig:simulambdavsK}(Color online) Dynamics of the quasiperiodic
kicked rotor in the vicinity of the critical regime. The rescaled
quantity $\ln\Lambda(K,t)$ vs. $K$ is plotted from $t=30$ to $t=40000$.
All curves intersect, to a very good approximation,
at a single point $(K_{c}\simeq6.4,\ln\Lambda_{c}\simeq1.6)$. This
multiple crossing indicates the occurrence of the metal-insulator
transition. Small deviations from crossing are due to the existence
of an irrelevant scaling parameter at finite time and residual correlations
in the disordered potential (see text). $K$ and $\epsilon$ are swept
along the straight line drawn in Fig.~\ref{fig:K-Epsilon}. Parameters
are $\kbar=2.85$, $\omega_{2}=2\pi\sqrt{5}$, $\omega_{3}=2\pi\sqrt{13}$.}

\end{figure}

A remarkable feature of Eq.~(\ref{eqlinscaling}) is that when $\ln\Lambda$
is plotted against $K$, the curves for different times $t$ should
intersect at a common point $(K_{c},\ln\Lambda_{c})$; and this crossing,
indicates the occurrence of the metal-insulator transition. This is
clearly visible in Fig.~\ref{fig:simulambdavsK}. Another interesting
feature of Eq.~(\ref{eqlinscaling}) is that the critical exponent
$\nu$ can be determined from the slope of $\ln\Lambda$ at $K_{c}$:
\begin{equation}
(\ln\Lambda)^{\prime}(K_{c},t)\equiv\left.\frac{\partial\ln\Lambda(K,t)}{\partial K}\right|_{K=K_{c}}\propto t^{1/3\nu}\;.\label{eq:lambdaprime}\end{equation}
This is the simplest procedure to evaluate the critical exponent:
$(\ln\Lambda)^{\prime}(K_{c},t)$ is evaluated by linear regression
of $\ln\Lambda$ vs $K$ in a small interval near $K_{c}$, giving
an exponent $\nu\simeq1.61\pm0.10$  (see Fig.~ref{fig:simulambdaprimevslnt}). 
The linear regime has nevertheless
very small size: $(K-K_{c})t^{1/3\nu}\ll1$, and neglecting non-linear
corrections lead to systematic errors on the estimation of $\nu$.
This is why the error $\pm0.1$ refers to systematic errors and not
to the uncertainty on the fitting parameters, which is much smaller
as easily seen in Fig.~\ref{fig:simulambdaprimevslnt}.

\begin{figure}
\begin{centering}
\psfrag{logt}{$\ln t$ } 
\psfrag{lambdaprime}{$(\ln\Lambda)'(K_{c})$} 
\psfrag{b}{ } 
\includegraphics[width=7cm]{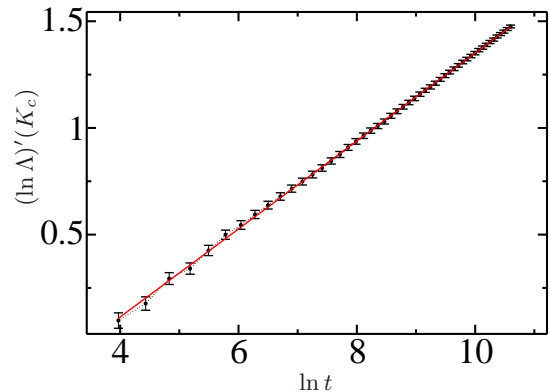} 
\end{centering}

\caption{\label{fig:simulambdaprimevslnt}(Color online) Linear regression
of $(\ln\Lambda)^{\prime}(K_{c},t),$ Eq.~(\ref{eq:lambdaprime})
vs. $\ln t$ for $t=30$ to $t=40000$ permits to extract the critical
exponent $\nu$ from the slope $1/3\nu$, which is $\nu=1.61$. It
is difficult to assess the uncertainty associated with this measurement
as it depends crucially on the interval of $K$ where the behavior
of $\ln\Lambda$ vs $K$ can be assumed to be {}``linear''. 
The parameters are the
same as in Fig.~\ref{fig:simulambdavsK}.}

\end{figure}

In practice, there are small systematic deviations from
Eq.~(\ref{eqlinscaling}). Such deviations can have different sources:
\begin{itemize}
\item the presence of an irrelevant scaling variable, that is when,
in addition to $(K-K_c)t^{1/3\nu}$, $(\ln\Lambda)$ depends also
on another scaling variable which vanishes in the limit $t\to \infty,$ but still plays a role
at short time;  
\item non-linear
dependence of the scaling variables in the stochasticity parameter
$K$; 
\item resonances due to the periods being well approximated
by a ratio of small integers. 
\end{itemize}
The latter one is specific
to our three-frequency dynamical system, but the former two also play
an important role in the standard Anderson model \cite{Slevin:PRL99,MacKinnon:JPC94}.
These small corrections can be taken into account 
-- following the method devised in \cite{Slevin:PRL99} for the Anderson model  -- 
by slightly modifying the basic scaling law, 
Eq.~(\ref{eq:scalinglawassumption}), in two ways:
introduce a non-linear of the argument of the $\mathcal{F}$ function
with $K-K_c$ in Eq.~(\ref{eq:scalinglnlambda}) on the one hand, and allow
to subtract irrelevant scaling corrections to $(\ln\Lambda)$
on the other hand. 
To minimize the the effect of resonances, we only retain data for 
sufficiently long
times and average over different initial
conditions, i.e. different quasi-momenta $\beta$ and phases $\varphi_{2}$
and $\varphi_{3}.$ 

We computed $\ln\Lambda$ for times up to $t=10^{6}$ kicks with an
accuracy of $0.15\%$, for which more than $1000$
initial conditions are required.
We analyze data over the full range of times
$t\in\left[10^{3},10^{6}\right]:$ The best fit is determined by minimizing the deviation
\begin{equation}
\chi^{2}=\sum_{K,t}\left[\frac{\ln\Lambda(K,t)-\mathcal{F}(K,t)}{\sigma(K,t)}\right]^{2}\;,\label{eq:lambdachi2}\end{equation}
where $\sigma(K,t)$ is the numerical uncertainty (one standard deviation) 
of the computed quantities $\ln\Lambda(K,t).$

In Fig.~\ref{fig:scalinglambdavsxsit13}, we plot the scaling function
corrected from the irrelevant scaling variable, 
as a function of $\xi(K)/t^{1/3}$. 
All data collapse almost
perfectly on the scaling function deduced from the model.

\begin{figure}
\begin{centering}
\psfrag{x}{$\ln\left(\xi/t^{1/3}\right)$ } \psfrag{y}{$\ln\Lambda_{s}$
} 
\includegraphics[width=8cm]{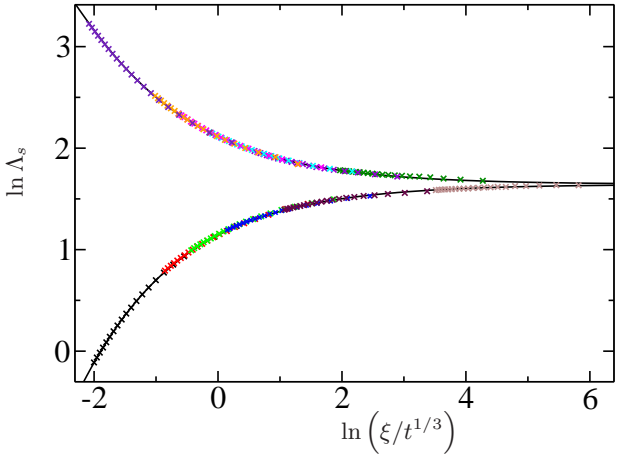} 
\end{centering}

\caption{\label{fig:scalinglambdavsxsit13} (Color online) $\ln\Lambda_{s}$,
after subtraction of
corrections due to the irrelevant scaling variable, plotted versus
$\ln\left(\xi/t^{1/3}\right)$
and the scaling function deduced from the model 
(black curve). The parameters are that
of the set $\mathcal{D}$ (see Table~\ref{Tab:set-of-parameters}).
The best fit estimates of the critical stochasticity and the critical
exponent are in this case: $K_{c}=8.09\pm0.01,\ \ln\Lambda_{c}=1.64\pm0.03$
and $\nu=1.59\pm0.01$.}

\end{figure}

 Since the measurement
errors in the data introduce some uncertainty in the determination
of the fitted parameters, the confidence intervals for the fitted
parameters were estimated using the bootstrap method which yields
Monte-Carlo estimates of the errors in the fitted parameters \cite{NumRec:92}.
The fitted parameters presented below are given with
the corresponding $68.2\%$ confidence intervals (standard errors).

\subsection{Universality of the critical exponent}

\label{sec:universality}

\begin{table}[t]

\begin{centering}
\begin{tabular}{c||c|c|c|c|c}
 & $\kbar$  & $\omega_{2}$  & $\omega_{3}$  & $K$  & $\varepsilon$ \tabularnewline
\hline
\hline 
$\mathcal{A}$  & $2.85$  & $2\pi\sqrt{5}$  & $2\pi\sqrt{13}$  & $6.24\rightarrow6.58$  & $0.413\rightarrow0.462$ \tabularnewline
\hline 
$\mathcal{B}$  & $2.85$  & $2\pi\sqrt{7}$  & $2\pi\sqrt{17}$  & $5.49\rightarrow5.57$  & $0.499\rightarrow0.514$\tabularnewline
\hline 
$\mathcal{C}$  & $2.2516$  & $1/\eta$  & $1/\eta^{2}$  & $4.98\rightarrow5.05$  & $0.423\rightarrow0.436$ \tabularnewline
\hline 
$\mathcal{D}$  & $3.5399$  & $\kbar/\eta$  & $\kbar/\eta^{2}$  & $7.9\rightarrow8.3$  & $0.425\rightarrow0.485$ \tabularnewline
\end{tabular}
\par\end{centering}

\caption{\label{Tab:set-of-parameters} The four sets of parameters considered:
$\kbar$, $\omega_{2}$ and $\omega_{3}$ control the microscopic
details of the disorder, while $\epsilon$ drives the anisotropy of
the hopping amplitudes.}

\end{table}

A key property of the Anderson transition is that 
the critical behavior can be described \cite{Slevin:PRL97,Efetov:AP83}
in a framework of universality classes. This means that the critical
behavior should not be sensitive of the microscopic details but should
depend only on the underlying symmetries of the system (e.g. time-reversal
symmetry). Irrelevant parameters  become negligible for sufficiently long times/large system
size, whereas the relevant parameter behavior is universal.
This brings the universality of the critical exponents. When considering
a system with pseudo-random disorder such as the quasi-periodic kicked
rotor, one could ask whether the universality is broken or not due
to correlations in the disorder potential. To answer the question,
we changed some parameters that govern the microscopic details of
the disorder potential of the quasi-periodic kicked rotor, namely
$\kbar$, $\omega_{2}$ and $\omega_{3}$ and the path along which
we cross the transition.

The computer time required in those sophisticated numerical studies
is very long. Therefore we chose to restrict ourselves to the \textit{detailed}
study of only four different cases, see Table \ref{Tab:set-of-parameters}.

\begin{table}
\begin{centering}
\begin{tabular}{c||c|c|c|c}
 & $K_{c}$  & $\ln\Lambda_{c}$  & $\mathbf{\nu}$  & $y$ \tabularnewline
\hline
\hline 
$\mathcal{A}$  & $6.36\pm0.02$  & $1.60\pm0.04$  & $\mathbf{1.58\pm0.01}$  & $0.71\pm0.28$ \tabularnewline
\hline 
$\mathcal{B}$  & $5.53\pm0.03$  & $1.08\pm0.09$  & $\mathbf{1.60\pm0.03}$  & $0.33\pm0.30$ \tabularnewline
\hline 
$\mathcal{C}$  & $5.00\pm0.03$  & $1.19\pm0.15$  & $\mathbf{1.60\pm0.02}$  & $0.23\pm0.29$ \tabularnewline
\hline 
$\mathcal{D}$  & $8.09\pm0.01$  & $1.64\pm0.03$  & $\mathbf{1.59\pm0.01}$  & $0.43\pm0.23$ \tabularnewline
\end{tabular}
\par\end{centering}

\caption{\label{Tab:critical-parameters} Best fit estimates of the critical
parameters $K_{c}$ and $\ln\Lambda_{c}$, the critical exponent $\nu$
together with their uncertainty (one standard deviation). $\nu$ is
expected to be universal whereas $\ln\Lambda_{c}$ and $K_{c}$ do
depend on anisotropy \cite{Evangelou:PRB94} and $\kbar$, $\omega_{2}$
and $\omega_{3}$. Irrelevant parameters are sensitive to microscopic
details, therefore $y$ is strictly positive and not universal.}

\end{table}

The estimated critical
parameters and their confidence intervals are given in 
Table \ref{Tab:critical-parameters}.
A \textit{typical} scaling function is drawn in 
Fig.~\ref{fig:scalinglambdavsxsit13}.

The most important point to be drawn from Table \ref{Tab:critical-parameters}
is that the estimates of the exponent $\nu$ for the four different
sets are in almost perfect agreement with each other and with the
estimate of $\nu$ based on numerical studies of the true random Anderson
model $\nu=1.57\pm0.02$ of the orthogonal symmetry class \cite{Slevin:PRL99}.
Note also that in the case of the quasiperiodic kicked rotor, the
critical stochasticity $K_{c}$ and $\ln\Lambda_{c}$ depend on: (i)
the anisotropy governed by the parameter $\varepsilon$ and (ii) $\kbar$,
$\omega_{2}$ and $\omega_{3}$.  
The dependence (i) of the critical disorder and critical
$\ln\Lambda$ on anisotropy is a typical feature of the Anderson transition
in anisotropic solids \cite{Soukoulis:PRB89,Evangelou:PRB94,Soukoulis:PRL96}.
The quasiperiodic kicked rotor may indeed be seen to correspond to
a model of random chains (coupled by terms scaling like $\varepsilon$
in the two transverse directions) considered in \cite{Evangelou:PRB94},
see Eq.~(\ref{eq:hoppingamplitudes3DQKR}). The dependence (ii) follows
from the relation between the initial {}``classical'' diffusion
constant (see section \ref{sec:KRStandard}) and the parameters $\kbar$,
$\omega_{2}$ and $\omega_{3}$. Such a
dependence was observed both numerically and experimentally for the
standard kicked rotor \cite{Shepelyansky:PD87,Steck:PRE2000}, and
was accounted for in terms of correlations between the kicks by Shepelyansky
in his early work \cite{Shepelyansky:PD87}.

The Anderson transition with the quasiperiodic kicked rotor is a robust
feature: we observed that, for certain mutually incommensurate
triplets ($\kbar$, $\omega_{2}$, $\omega_{3}$), systematic deviations
to scaling (such as resonances) can occur for intermediate times,
but eventually vanish.

\section{Conclusion\label{sec:Conclusion}}

We discussed in detail in the present work the first unambiguous
evidence of the Anderson transition in 3D with atomic matter waves
with atomic matter waves by realizing experimentally a quasiperiodic
kicked rotor. This allowed us to put into evidence the existence of
the transition and to measure its critical exponent thanks to a finite-time
scaling procedure. Our numerical result $\nu=1.59\pm0.01$ is in perfect
agreement with the current value for the Anderson model, and is compatible
with our experimental determination $1.4\pm0.3$. We have also shown that
the quasiperiodic kicked rotor exhibits the same critical phenomena
as the truly random Anderson model, and therefore that both systems belong to the
same (orthogonal) universality class.

These results are particularly relevant since they
show that it is possible to explore a system like the Anderson model,
that played an important hole in many areas of physics but resisted
thorough experimental investigations. One can guess that this kind
of analogy will be extended to other models in the near future, as
evidenced by the work of Wang and Gong \cite{Gong:HarperKR:PRA08}
concerning the analogy of a quantum kicked rotor and the Harper model.
This shall open new and exciting tracks in cold-atom physics. These
analog models can even prove more flexible and more powerful than
the original ones, as, for example, our Anderson-equivalent system
can very easily be extended to higher dimensions by introducing new
incommensurate frequencies.  Intermediate situations like a 2D kicked rotor with two 
or three incommensurate frequencies might be a convenient solution from the experimental point of view. 
This can hardly be done
in condensed-matter systems or even in the ultracold atom realization
of the 1D Anderson model \cite{Bouyer:AndersonBEC:N08}. The theoretical
study of quantum phase transitions in high dimensions will most probably
be boosted as experimental results become available. We are presently
working in this direction: Numerical and experimental determinations
of the critical exponents in four dimensions seems feasible.

\begin{acknowledgments}
The authors are pleased to acknowledge H. Lignier for fruitful discussions.
\end{acknowledgments}

%\bibliographystyle{apsrev}
%\bibliography{longAnderson-V11,ArtDataBase}

\begin{thebibliography}{81}
\bibitem[{not({\natexlab{a}})}]{note:quasi-momentum}
\bibinfo{note}{When the quasi-momentum is taken into account, $m$ should simply
  be replaced by $m+\beta$, leaving the conclusions unaffected.}
\bibitem[{not({\natexlab{b}})}]{note:hopping}
\bibinfo{note}{$W(\theta)$ is singular for $K/\kbar > \pi$ and the behavior of
  $W_r$ is complicated. However, this non-physical singularity can be avoided
  by a mapping on a different tight-binding model \cite{Shepelyansky:PRL86}.}
\end{thebibliography}

\end{document}